\documentclass[lettersize,journal]{IEEEtran}
\usepackage{amsmath,amsfonts}
\usepackage{array}
\usepackage[caption=false,font=normalsize,labelfont=sf,textfont=sf]{subfig}
\usepackage{textcomp}
\usepackage{stfloats}
\usepackage{url}
\usepackage{verbatim}
\usepackage{graphicx}
\usepackage{cite}
\usepackage{caption}
\usepackage{multirow}
\usepackage[linesnumbered,lined,boxed,commentsnumbered,ruled,longend, vlined]{algorithm2e}

\newcommand{\T}{{\mathcal T}}
\newcommand{\cO}{{\mathcal O}}

\usepackage{booktabs} 
\usepackage{enumitem}
\usepackage{mathtools}
\usepackage{todonotes}
 \usepackage{nicematrix}

 \usepackage{mathtools, nccmath}

\newtheorem{definition}{Definition}
\newtheorem{example}{Example}

\newtheorem{lemma}{Lemma}
\newtheorem{theorem}{Theorem}

\newtheorem{corollary}{Corollary}

\newcommand{\ra}[1]{{\color{black}#1}}
\newcommand{\new}[1]{{\color{black}#1}}

\begin{document}

\title{Fast Transaction Scheduling in Blockchain Sharding}

\author{Ramesh Adhikari,
Costas Busch,
Miroslav Popovic,~\IEEEmembership{Member,~IEEE}

\thanks{Ramesh Adhikari and Costas Busch are with the School of Computer and Cyber Sciences, Augusta University, Augusta, GA, USA (emails: RADHIKARI@augusta.edu; KBUSCH@augusta.edu).}%
    \thanks{Miroslav Popovic is with the Faculty of Technical Sciences, University of Novi Sad, Novi Sad, Serbia (email: miroslav.popovic@rt-rk.uns.ac.rs).}%
    
}



\maketitle

\begin{abstract}
Sharding is a promising technique for addressing the scalability issues of blockchain, and this technique is especially important for IoT, edge, or mobile computing. It divides the $n$ participating nodes into $s$ disjoint groups called shards, where each shard processes transactions in parallel.
We examine batch scheduling problems on the shard graph $G_s$, where we find efficient schedules for a set of transactions. 
First, we present a centralized scheduler where one of the shards is considered as a leader, who receives the transaction information from all of the other shards and determines the schedule to process the transactions. 
For general graphs, where a transaction and its accessing objects are arbitrarily far from each other with a maximum distance $d$, the centralized scheduler provides $O(kd)$ approximation to the optimal schedule, where $k$ is the maximum number of shards each transaction accesses.
Consequently, for a Clique graph
where shards are at a unit distance from each other,
we obtain $O(k)$ approximation to the optimal schedule.
Next, we provide a centralized scheduler with a bucketing approach that offers improved bounds for the case where $G_s$ is a line graph,
or the $k$ objects are randomly selected. 
Finally, we provide a distributed scheduler where shards do not require global transaction information.
We achieve this by using a hierarchical clustering of the shards and using the centralized scheduler in each cluster. 
We show that the distributed scheduler has a competitive ratio of $O(\mathcal{A_\mathcal{CS}} \cdot \log d \cdot \log s)$, where $\mathcal{A_\mathcal{CS}}$ is the approximation ratio of the centralized scheduler.  
To our knowledge, we are the first to give provably fast transaction scheduling algorithms for blockchain sharding systems. We also present simulation results for our schedulers and compare their performance with a lock-based approach. The results show that our schedulers are generally better with up to 3x lower latency and 2x higher throughput.
\end{abstract}


\section{Introduction}
\label{sec:introduction}

Blockchains are known for their special features such as fault tolerance, transparency, non-repudiation, immutability, and security \cite{survey-of-onsensus}. Thus, blockchains have been used in various domains, such as cryptocurrency \cite{bitcoin, ethereum}, healthcare \cite{mcghin2019blockchain}, insurance \cite{10575641}, digital forensics \cite{akbarfam2023forensiblock,akbarfam2023deep}, supply chain management \cite{azzi2019power}, edge and scientific computing \cite{al2021scichain}.
\ra{Blockchain technology is also gaining interest from Transportation Systems\cite{9430722}, Mobile Edge Computing\cite{9387145}, cloud-edge-end and cooperative network~\cite{yuan2022coopedge}, 
because of its secure, decentralized, and fault tolerance attributes~\cite{9430722,9387145,yuan2022coopedge}.} 
 However, the primary drawback of blockchain is its scalability and its performance, which is dependent on the throughput of the transactions \cite{cocco2017banking,chen2020blockchain}. Thus, there is a research gap to address the scalability issues of blockchain.
 
To append a new block in the blockchain network, each participating node needs to reach a consensus, which is a time and energy-consuming process \cite{adhikari2023lockless, adhikari2024spaastable}. Moreover, each node is required to process and store all transactions, which leads to scalability issues in the blockchain system. 
To improve scalability and performance, {\em sharding} protocols have been proposed, such as Elastico \cite{Elastico}, OmniLedger \cite{OmniLedger}, RapidChain~\cite{Rapidchain}, Sharper \cite{amiri2021sharper}, X-Shard\cite{xu2024x} and ByShard \cite{Byshard}. However, these sharding protocols do not provide formal analysis for the schedule time complexity (i.e. how fast the transactions are processed). Sharding divides participating nodes into smaller groups called shards that allow the parallel processing of transactions. In the sharded blockchain, independent transactions are processed and committed in multiple shards concurrently, which improves the blockchain system's throughput.

\begin{figure}[!ht]
  
  \centering
  \includegraphics[width=0.45\textwidth]{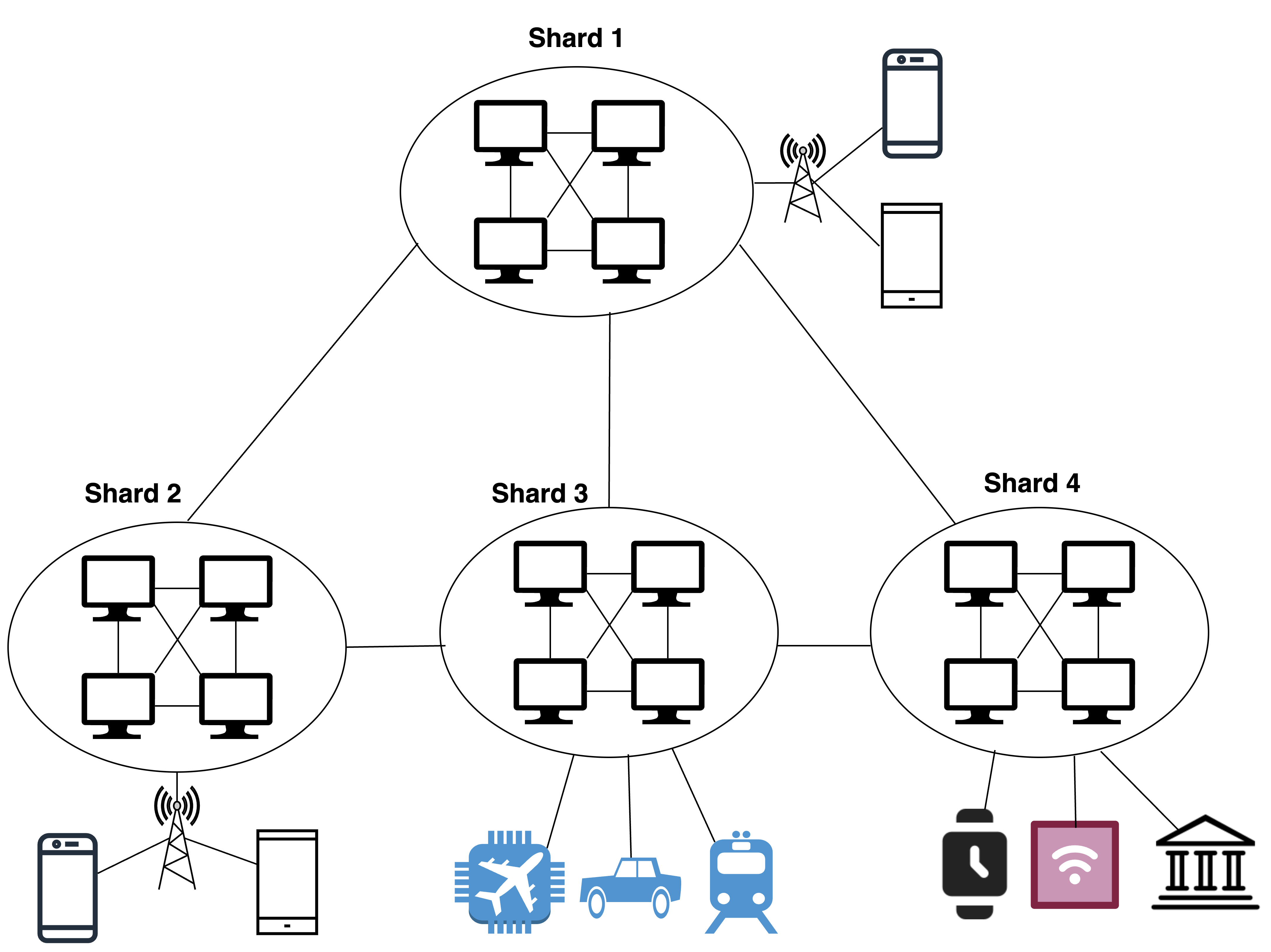}
  \caption{Edge, IoT and Mobile devices with blockchain sharding.}
  \label{fig:iot_with_sharding}
\end{figure}

\ra {Mobile-edge computing data \cite{9387145} and cloud-edge data \cite{yuan2022coopedge} can be processed with a blockchain that utilizes sharding, which reduces storage, communication, and computation costs~\cite{Rapidchain}. Figure \ref{fig:iot_with_sharding} shows sharding with IoT and mobile devices, where each shard operates on edge computing servers, connects with IoT devices, and handles transaction execution.
In Figure \ref{fig:iot_with_sharding}, Shards $1$ and $2$ are responsible for processing and handling mobile data, Shard $3$ stores data related to transportation and connected with devices such as airlines, cars, and trains, and Shard $4$ stores data related to IoT devices, connecting with devices like smartwatches, Wi-Fi routers, and banks. Transactions can be generated in any shard, and the shard is responsible for processing a particular transaction.
We assume that these devices generate transactions and send them to the shards for storing and processing. Our goal is to process/store these transactions as fast as possible. Our scheduler determines the order in which these transactions are processed in a conflict-free manner, as some transactions might conflict by attempting to store or update the same object (e.g., entity or account).
}

We consider a blockchain system consisting of $n$ nodes, which are further divided into $s$ shards, and each shard consists of a set of nodes.
Shards are connected in a graph network $G_s$ with a diameter $D$, and each shard holds a subset of the objects (accounts).
We assume that transactions are distributed across the shards, and each transaction accesses at most $k$ accounts.
A transaction $T$ initially is in one of the shards, which is called the {\em home shard} for $T$. 
Similar to other sharding systems~\cite{Byshard, adhikari2023lockless,adhikari2024spaastable}, each transaction $T$ is split into subtransactions,
where each subtransaction accesses an account. 
A subtransaction of $T$ is sent to the {\em destination shard} that holds the respective account. The maximum distance between the home shard of a transaction and the respective destination shards in $G_s$ is at most $d\leq D$. In other words, the transaction and its accessing shards (objects) are arbitrarily far from each other with a maximum distance of $d$. 

All home shards process transactions concurrently.
A problem occurs when {\em conflicting} transactions
try to access the same account simultaneously.
In such a case, the conflict prohibits the transactions to be committed concurrently
and forces them to serialize \cite{ adhikari2024spaastable}.
Our proposed {\em scheduling algorithms}
coordinate the home shards and destination shards to process the transactions (and respective subtransactions) in a conflict-free manner.
Each destination shard maintains a local blockchain of the subtransactions that are sent to it.
The global blockchain can be constructed (if needed) 
by combining the 
local blockchains at the shards~\cite{adhikari2023lockless}.

We consider batch problem instances where a set of transactions $\T$ \ra{are generated by (IoT and mobile) devices that} reside at shards. 
We study the shard model in different graph topologies such as General graph, Clique (Complete) graph, and Line graphs and provide the analysis for execution time. 
\ra{Considering the different graph topologies for the IoT and mobile computing is important as presented in\cite{de2018impact, mamat2019network}}.

   The scheduler determines the time step for each transaction $T_i \in \T$ to process and commit. The execution of our scheduling algorithm is synchronous, where the timeline is divided into time steps referred to as {\em rounds}. 
Similar to the work in~\cite{adhikari2024spaastable}, we consider that the duration of a round is sufficient to allow 
the execution of the PBFT consensus algorithm \cite{PBFT} in each shard.
A round is also the time to send a message between shards in a unit distance.
The main goal of a scheduling algorithm is to efficiently and fairly process all transactions while minimizing the total execution time (makespan) until all transactions are either committed or aborted. 
Unlike previous sharding approaches~\cite{Byshard,OmniLedger,Rapidchain}, our scheduling algorithms do not require locking mechanisms for concurrency control.
Hence, we can also say our algorithms are lock-free.

\renewcommand{\arraystretch}{1.4}
 \begin{table*}[t]
\centering
\small
\begin{tabular}{|l|l|l|l|l|l|l|}

\hline
 \multirow{2}{*}{{  Shards are Connected  as}}  & \multicolumn{2}{l|}{ { Approximation to Optimal Schedule of}} \\ \cline{2-3} 
    &{ Centralized Scheduler}  & { Distributed Scheduler}  \\ \hline

 General Graph & $O(kd)$ & $O(kd\cdot \log d \cdot \log s)$      \\ \cline{1-3}
  \multirow{1}{*}{Clique Graph with unit distance} & $O(k)$ &$O(k)$ [same with centralized]  \\ \hline
 Hypercube, Butterfly, and $g$-dimensional Grid  Graph & $O(k \log s)$ & $O(k \cdot \log d \cdot \log^2 s)$     \\ \cline{1-3} 

   General Graph where transaction access $k$ objects randomly  &$O(k  (k+\log s) \log D)$ & $O(k (k + \log s)\log D  \cdot \log d \cdot \log s)$
 \\ \cline{1-3} 
  Line Graph   &  $O(k \sqrt{d} \log D)$& $O(k\sqrt{d} \log D \cdot \log d \cdot \log s)$     \\ \cline{1-3} 
 
 \hline

\end{tabular}
\renewcommand{\arraystretch}{1}
\caption{Summary of the proposed centralized and distributed scheduler approximation to the optimal schedule. The used notations are as follows: $s$ represents the total number of shards, $k$ denotes the maximum number of shards (objects) accessed by each transaction, $d$ denotes the worst distance between any transaction (home shard) and its accessing objects (destination shard), $D$ denotes the diameter of shard graph $G_s$. 
}
\label{tbl:contribution-summary}
\end{table*}

\subsection{Contributions}
To our knowledge, we give the first provably efficient batch transaction scheduling algorithms for blockchain sharding systems. 
Table \ref{tbl:contribution-summary} has a summary of our contributions which we also described as follows: 
\begin{itemize}

\item {\bf Centralized Scheduler:} First, we provide a Centralized scheduler where one of the shards is considered as a leader, and all of the other shards send their transaction information to the leader, and the leader determines the schedule to process the transactions.
Consider a general graph $G_s$ with diameter $D$. Suppose that each transaction accesses at most $k$ objects,
where the distance between a transaction and its accessing objects is at most $d \leq D$.
The centralized scheduler provides an $O(kd)$ approximation to the optimal schedule.
Hence, when $G_s$ is a Clique graph,
where shards are at a unit distance from each other (thus, $d = 1$),
we obtain an $O(k)$ approximation to the optimal schedule.
\new {Moreover, we obtain an $O(k \log s)$ approximation for the Hypercube, Butterfly, and $g$-dimensional Grid where $g=O(\log s)$, since in all these cases $d = O(\log s)$.}

\item {\bf Centralized Scheduler with Buckets:} Next, we give the centralized scheduler with a bucketing approach. 
We divide the transactions into buckets $B_0, B_1, \ldots$,
according to the maximum distance between the home shard of the transaction and its accessing objects (i.e., destination shards).
$B_i$ consists of the transactions whose maximum distance between 
the home shard and its destination shard is in range $[2^i, 2^{i+1})$. This approach offers improved bounds for special cases:
for the General graph where the $k$ objects are chosen randomly per transaction we give $O(k(k+\log s)\log D)$ approximation, and
for the Line graph, we provide $O(k\sqrt{d} \log D)$ approximation to the optimal schedule. 


\item {\bf Distributed Scheduler:} Finally, we give a distributed scheduler where shards are not aware of global transaction information. We use a hierarchical clustering approach \cite{gupta2006oblivious} for the shard network so that independent transactions in different clusters can be scheduled and committed concurrently. The basic idea is to invoke the centralized scheduler within each cluster of the hierarchy to make a distributed scheduler. We show that the distributed scheduler has a competitive ratio of $O(\mathcal{A_\mathcal{CS}} \cdot \log d \cdot \log s)$, where $\mathcal{A_\mathcal{CS}}$ is the approximation ratio of the centralized scheduler.

\ra{\item {\bf Simulation Results:} \new{ We simulate our proposed transaction scheduling algorithms and analyze their average throughput, latency, and number of message exchanges. The simulation results show that our schedulers are better than a lock-based (Byshard \cite{Byshard}) approach by achieving up to $2$x higher throughput, $3$x lower latency, and $6$x fewer message exchanges.}}

\end{itemize}

{\bf Paper Organization:}
The rest of this paper is structured as follows: Section \ref{sec:related-work} provides related works. Section \ref{preliminaries} describes the preliminaries for this study and the sharding model. 
Section~\ref{sec:centralized-scheduler} presents a centralized scheduling algorithm. In Section \ref{sec:centralized-scheduler-with-bucket}, we provide a centralized scheduler with the bucketing approach.
Section~\ref{sec:fully-distributed} generalizes the techniques to a fully distributed setting. In Section \ref{sec:simulation-results}, we provide the simulation results of each proposed algorithm. 
Finally, we give our conclusions in Section~\ref{sec:conclusion}.

\section{Related Work}
\label{sec:related-work}

Extensive research has been proposed to address the scalability issues of blockchain in the consensus layer~\cite{yin2019hotstuff,Jalal-Window,jalalzai2019proteus,jalalzai2021hermes,stathakopoulou2019mir,spiegelman2022bullshark,danezis2022narwhal}. While these proposed algorithms and protocols have made some improvements in scalability, the system's performance still degrades as the network size increases.
To tackle the scalability issue of blockchain, various sharding protocols \cite{Elastico, OmniLedger,Rapidchain,Byshard,dang2019towards,amiri2021sharper,adhikari2023lockless,li2023lb} have been proposed. These protocols have shown promising enhancements in the transaction throughput of blockchain by processing non-conflicting transactions in parallel in multiple shards. However, none of these protocols have specifically explored the batch scheduling of transactions within a sharding environment, and they do not provide an extensive theoretical analysis for transaction scheduling.
\ra{There are many research work has been done to securely store and process the IoT and mobile computing data in blockchain\cite{queralta2021blockchain, li2021blockchain, 9387145}, and blockchain sharding \cite{9324984,ren2023data}.  However, these works also lack the theoretical analysis for the optimal schedule.}

To process transactions parallelly in the sharding model, some research work \cite{OmniLedger,Byshard,set2022service} has used two-phase locking for concurrency control. However, locks are expensive because when one process locks shared data for reading/writing, all other processes attempting to access the same data set are blocked until the lock is released, which lowers system throughput. Moreover, locks if not handled and released properly may cause deadlocks. Our scheduling algorithm does not use locks, as concurrency control is managed by the scheduler by processing non-conflicting transactions concurrently. In~\cite{adhikari2023lockless} the authors propose lockless blockchain sharding using multi-version concurrency control. However, they lack an analysis and technique for optimal scheduling. Moreover, they do not provide the benefits of locality and optimization techniques for transaction scheduling.

Several works have been conducted on transaction scheduling in shared memory multi-core systems, distributed systems, and transactional memory. In a recent work \cite{busch2023stable}, a stable scheduling algorithm was proposed specifically for software transactional memory systems, considering transactions generated by an adversarial model. Similarly, in research works \cite{attiya2015directory, sharma2014distributed,sharma2015load}, authors explored transaction scheduling in distributed transactional memory systems aimed at achieving better performance bounds with low communication costs.  Authors in work \cite{busch2017fast} provided offline scheduling for transactional memory, where each transaction attempts to access an object, and once it obtains the object, it executes the transaction. In another work \cite{busch2022dynamic}, the authors extended their analysis from offline to online scheduling for the transactional memory model. However, these works do not address transaction scheduling problems in the context of blockchain sharding. This is because, in the transactional memory model, the object is mobile, and once the transaction obtains the object, it immediately executes the transaction. In contrast, in blockchain sharding, the object is static in the shard, and there is a confirmation scheme to confirm and commit each subtransaction consistently in the respective accessing shard.

In a recent work \cite{ adhikari2024spaastable}, the authors provide a stability analysis of blockchain sharding considering adversarial transaction generation. However, their solution does not consider different types of sharding graphs, and they also do not provide a theoretical analysis of the optimal approximation for the scheduling algorithm.


\section{Technical Preliminaries}
\label{preliminaries}

A block is a data structure that consists of a set of transactions. The block header contains additional metadata, including the block hash, previous block hash, block sequence, etc.
Blockchain is simply a chain of blocks, where a block points to its predecessor (parent) through its hash, which makes blocks immutable. 
A blockchain is essentially implemented as a decentralized peer-to-peer ledger where the blockchain is replicated across multiple interconnected nodes.


\paragraph{\bf Blockchain Sharding Model}
 We consider  blockchain sharding model which is similar to ~\cite{adhikari2023lockless,Byshard, adhikari2024spaastable}, consisting of $n$ nodes 
which are 
partitioned into $s$ shards $S_1, S_2,\dots, S_s$ such that $S_i \subseteq \{1, \ldots, n\}$, for $i \neq j$, $S_i \cap S_j = \emptyset$, $n = \sum_i |S_i|$, and $n_i = |S_i|$ denotes the number of nodes in shard $S_i$.

Let $G_s = (V,E,w)$ denote a weighted graph of shards, where $V = \{S_1, S_2,\dots, S_s\}$, and the edges $E$ correspond to the connections between the shards,
and the weight represents the distance between the shards.
We assume that $G_s$ is a clique, namely, any two shards can communicate with each other. The communication delay between two shards is expressed by the edge weight connecting them.

Shards communicate with each other via message passing, and here, we are not focusing on optimizing the message size.
Moreover, all honest nodes in a shard agree on each message before transmission
(e.g. running the PBFT \cite{PBFT} consensus algorithm within the shard). Similar to~\cite{Byshard,hellings2022fault,adhikari2024spaastable}, we 
use a
{\em cluster-sending} protocol for reliable and secure communication between shards, satisfying the following properties:
    (1) Shard $S_x$ sends data $\mathfrak{D}$ to shard $S_y$ if there is an agreement among the honest nodes in $S_x$ to send $\mathfrak{D}$.
    (2) All honest nodes in recipient shard $S_y$ will receive the same data $\mathfrak{D}$.
    (3) All non-faulty nodes in sender shard $S_x$ receive confirmation of data $\mathfrak{D}$ receipt.

Similar to \cite{adhikari2024spaastable}, for secure inter-shard communication, we use a broadcast-based protocol that can operate in a system with Byzantine failures \cite{hellings2022fault}. For the communication between shards $S_1$ and $S_2$,  shard $S_1$ uses a consensus protocol to reach
agreement on a value $v$. Then, a set $M_1 \subseteq S_1$ of $f_1+1$  nodes in $S_1$ and a set $M_2 \subseteq S_2$ of  $f_2 + 1$ nodes in $S_2$ are chosen, where $f_i$ is the number of faulty nodes in shard $S_i$. Finally, each node in $M_1$ is instructed to broadcast $v$ to all nodes
in $M_2$. Due to the choices of $M_1$ and $M_2$, it is guaranteed that at least one non-faulty node in $S_1$ will send a value to
a non-faulty node in $S_2$, which is sufficient to receipt and confirmation of $v$ in $S_2$.

We assume communication is synchronous and time is divided into {\em rounds}.
A round is the time it takes to reach a consensus within a shard. A round is also the time to send a message from one shard to another at a unit distance.
The above cluster-sending properties between two shards at unit distance are guaranteed to be satisfied within a single communication round.
Moreover, the distance between two shards is measured as the number of rounds required to send a message from one shard to another over the graph $G_s$.

Each shard consists of a local blockchain (which is part of the global blockchain) according to the subtransactions it receives and commits. 
We use $f_i$ to represent the number of Byzantine nodes in shard $S_i$.
To guarantee consensus on the current state of the local blockchain, we assume that every shard executes the PBFT \cite{PBFT} consensus algorithm.
In order to achieve Byzantine fault tolerance, 
similar to previous works \cite{Byshard, adhikari2024spaastable,adhikari2023lockless}, we assume each shard $S_i$ consists of  $n_i > 3 f_i$ nodes, where $n_i$ is the total number of nodes in $S_i$. 



Suppose we have a set of shared accounts $\mathcal{O}$ (which we also call shared {\em objects}). 
Similar to previous works in \cite{Byshard, adhikari2023lockless, adhikari2024spaastable}, we assume that each shard is responsible for a subset of the shared objects (accounts).
Thus, $\mathcal{O}$ is partitioned into disjoint subsets $\mathcal{O}_1, \ldots, \mathcal{O}_s$, where $\mathcal{O}_i$ is the set of objects handled by shard $S_i$.


  


Let $D$ denote the diameter of graph $G_s$.
We assume that the distance between the transaction and its accessing objects ranges between $1$ to $d$, where $d\leq D$.

\paragraph{\bf Transactions and Subtransactions}
Let's consider a set of transactions $\T = \{ T_1, T_2, \ldots\}$ that are distributed across different shards. 
Let $S(T_i)$ denote the {\em home shard} of transaction $T_i$,
which is the shard responsible for processing $T_i$.
The home shard may receive $T_i$ from one of the IoT or mobile devices associated with the shard.
In this work, we consider batch transactions that are appear in the home shards and need to be processed.

As in \cite{Byshard, adhikari2023lockless,adhikari2024spaastable}, a transaction $T_i$ consists of subtransactions $T_{i,a_1},\ldots,T_{i,a_j}$,
such that each subtransaction $T_{i,a_l}$ accesses objects only in $\mathcal{O}{a_l}$ associated with shard $S{a_l}$. Thus, each subtransaction $T_{i,a_l}$ has a respective {\em destination shard} $S_{a_l}$. The home shard of $T_i$ sends subtransaction $T_{i,a_l}$ to shard $S_{a_l}$ for processing, where it is appended to the local blockchain of $S_{a_l}$. The subtransactions within a transaction $T_i$ are independent, meaning they do not conflict and can be processed concurrently. Like previous work in \cite{Byshard,adhikari2024spaastable}, each subtransaction $T_{i,a_l}$ consists of two parts: (i) a condition check, where it verifies whether a condition of the objects in $\mathcal{O}{a_l}$ is satisfied, and (ii) the main action, where it updates the values of the objects in $\mathcal{O}{a_l}$.
\begin{example}
Consider a transaction $T_1$ that involves read-write operations on accounts with certain conditions. For instance, let $T_1$ be ``Transfer 3000 from Ray's account to Alex's account if Ray has 8000, Alex has 500 and Balen has 700''. The home shard of $T_1$ divides this transaction into three subtransactions $T_{1,r}, T_{1,a}, T_{1,b}$, where the destination shards $S_{r}$, $S_{a}$, and $S_{b}$ handle Ray's, Alex's, and Balen's accounts respectively:
\end{example}

\begin{itemize}[itemindent=0.7cm]
\item[$T_{1,r}$] - condition: ``Check if Ray has 8000''
\item[] - action: ``Deduct 3000 from Ray's account''
\item[$T_{1,a}$] - condition: ``Check if Alex has 500''
\item[] - action: ``Add 3000 to Alex's account''
\item[$T_{1,b}$] - condition: ``Check if Balen has 700''
\end{itemize}
The home shard of $T_1$ sends the subtransactions to their respective destination shards. If the conditions are met and the transactions are valid, then the destination shards commit the subtransactions in the local blockchains, which implies the entire transaction $T_1$ implicitly commits as well. However, if any of the conditions are not met, they are not added to the local blockchains, resulting in the abortion of $T_1$ as well.

\begin{definition}[{\sc Conflict}]
 \label{def:conflict}
Transactions $T_i$ and $T_j$ are said to be {\em conflict} if they access the same object $o_z \in \mathcal{O}$ and at least one of these transactions 
writes (updates) the value of object $o_z$. 
 \end{definition}
 Transactions that conflict should be processed in a sequential manner
to guarantee atomic object update.
In such a case, their respective subtransactions 
should serialize in the exact same order 
in every involved shard.
 To resolve the conflict between two transactions $T_i$ and $T_j$ while accessing an object $o_z$, we schedule them one after another in such a way that $T_i$ executes before $T_j$ or vice versa.

\section{Centralized Scheduler}
\label{sec:centralized-scheduler}
We consider a shard graph $G_s$ built from $s$ shards, where a set of transactions $\T$ and a set of objects $\cO$ are distributed across the $G_s$. In this centralized scheduler, we consider one shard as the {\em leader shard $S_\ell$}, which is responsible to create a schedule for all the transactions.
The home shards of the various transactions send the transaction information to the leader shard $S_\ell$ which it then runs a greedy coloring algorithm to schedule the transactions in a conflict-free manner. 
The details appear in Algorithm \ref{alg:centralized-scheduler}.

Consider each transaction $T_i\in \T$ accesses at most $k$ objects from a set of objects $\cO$. Suppose $G_\T$ is the transactions interference (conflict) graph where each node represents a transaction and an edge between
two nodes corresponds to a conflict between them. Recall that a conflict occurs when
the respective transactions try to access the same account (object) $o_j\in \cO$ in the same shard.

Algorithm~\ref{alg:centralized-scheduler} uses two phases.
In Phase 1, it performs a greedy
vertex coloring of $G_{\T}$, which assigns execution times to each transaction based on
their respective colors. A valid coloring of $G_\T$ assigns a unique positive
integer to each transaction such that two adjacent transactions (conflicting transactions)
receive different colors. The colors correspond to the distinct time
steps (also called rounds) where the transactions are processed for commit. In Phase 2, our algorithm takes four steps for each assigned color to confirm and commit the transactions with that color in each destination shard.
We continue to describe each phase of the algorithm.

\begin{algorithm}[t]
\small
\caption{{\sc CentralizedScheduler} ($G_s$, $\T$, $S_\ell$)}
\label{alg:centralized-scheduler}
\SetKwInOut{Input}{Input}\SetKwInOut{Output}{Output}
\tcp{txn: transaction; txns: transactions}

\Input{Shard graph $G_s$;
set of txns $\T$ that need to be processed in $G_s$; $S_\ell$ is the leader shard for txns $\T$; 
}

\BlankLine
  \BlankLine

        \tcp{Phase 1: Graph Coloring and determining schedule}
         
        Home shard of each txn $T_i \in \T$ shares the txn information with the leader shard $S_\ell$\;
         
       Leader shard $S_\ell$ constructs $G_\T$, and runs a greedy coloring on $G_\T$ which returns $\xi$ colors;
       
       Leader shard $S_\ell$ informs the home shards of each txn $T_i \in \T$ about the respective txn colors\;

        \BlankLine
          \BlankLine
        \tcp{Phase 2: confirming and committing}
        \tcc{processed colored txn   at appropriate round(s)}
        \For{color $clr \leftarrow 1$ \KwTo $\xi$}{
            \tcc{Step 1: Home Shard}
           The home shard of txn $T_i \in \T$ of color $clr$, splits $T_i$ into subtransactions ($T_{i,j}$) and sends them to respective destination shards for voting\;
            
            \BlankLine
            \tcc{Step 2: Destination shards}
            At destination shard $S_j$, if $T_{i,j}$ is valid and its condition is satisfied, then $S_j$ sends {\em commit vote} for $T_{i,j}$ to home shard of $T_i$; otherwise, $S_j$ sends {\em abort vote} for $T_{i,j}$ to home shard of $T_i$\;
            
            \BlankLine
            \tcc{Step 3: Home Shard}
            The home shard of $T_i$ checks whether all commit votes are received for $T_i$ from its destination shards; if true, the home shard sends {\em confirm commit} to destination shards; otherwise (if any abort vote received), the home shard sends {\em confirm abort} to destination shards\;
            
            \BlankLine
            \tcc{Step 4: Destination Shards}
            At destination shard $S_j$ of $T_{i,j}$, if $S_j$ receives confirm commit for $T_{i,j}$  (from home shard of $T_i$), then $S_j$ commits $T_{i,j}$ by appending it in the local blockchain; 
            Otherwise, if $S_j$ receives confirmed abort (from home shard of $T_i$), then $S_j$ aborts $T_{i,j}$\;
        }
\end{algorithm}

{\bf Phase 1:} In this phase, the home shard of each transaction $T_i\in \T$ sends its transaction information to the leader shard $S_\ell$. Then, the leader shard constructs the transactions conflict graph $G_\T$ and runs the greedy coloring algorithm on $G_\T$. The greedy coloring algorithm assigns the colors to each transaction of $G_\T$ using at most $\xi$ colors. After that, the leader shard informs the home shard of each transaction $T_i \in \T$ about the respective transaction color.

{\bf Phase 2:} In this phase, each transaction is processed based on a computed schedule using the greedy coloring in Phase~1. Each home shard knows its transaction committing time and proceeds to determine whether it can be committed or not.
Each of the $\xi$ colors takes $4$ steps for the confirmation (i.e., $4$ communication interactions between home shard and destination shards).
So, for each color  $clr \in \xi$ in Step 1, each home shard of transaction $T_i \in \T$ of color $clr$ splits the transaction $T_i$ into subtransactions ($T_{i,j}$) and sends them to the respective destination shards parallelly to check whether that transaction (i.e. its subtransactions) can be committed or not. In Step 2, each destination shard $S_j$ checks the conditions and validity of the subtransactions $T_{i,j}$ (such as account balance) and sends a commit vote to the home shard of $T_i$ if the subtransaction $T_{i,j}$ is valid; otherwise sends an abort vote to home shard of $T_i$ if the subtransaction $T_{i,j}$ is invalid. In Step~3, the home shard of $T_i$ collects the votes, and if it receives all the commit votes for a transaction $T_i$ from its destination shards, then it sends the confirmed commit message to the respective destination shards. Otherwise, if the home shard of $T_i$ receives any abort vote, then it will send a confirmed abort message to all respective destination shards. In Step 4, each destination shard either commits the subtransactions $T_{i,j}$ and appends them to a local ledger or aborts the subtransactions according to the message received from the home shard of $T_i$.

\begin{figure*}[!ht]
  
  \centering
  \includegraphics[width=0.99\textwidth]{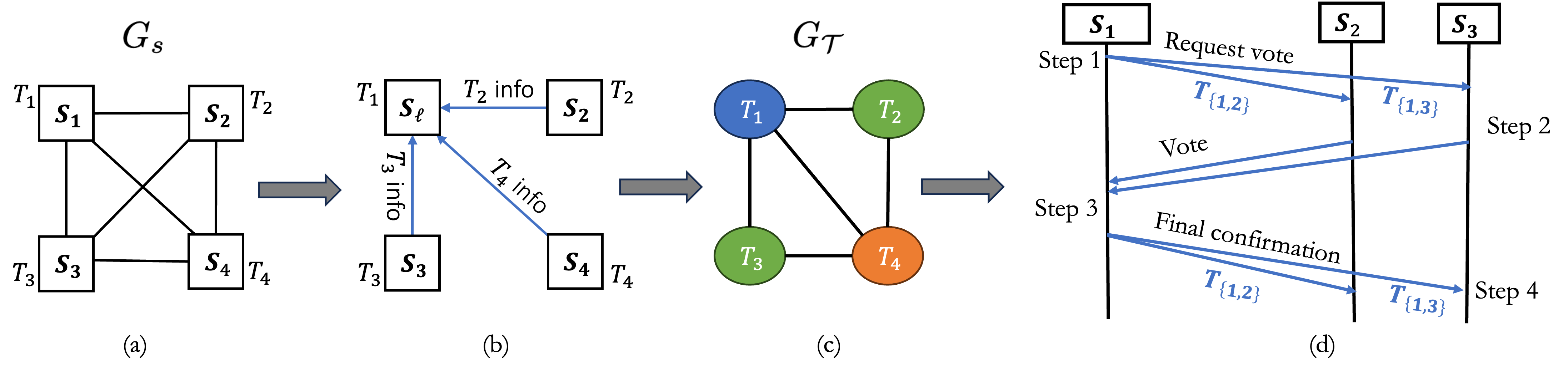}
  \caption{Simple representation of transaction processing using Algorithm \ref{alg:centralized-scheduler}, where (a) represents four shards, each having one pending transaction to process. Figure (b) and (c) represent Phase 1, and Figure (d) represents Phase 2 of the Algorithm~\ref{alg:centralized-scheduler}.}
  \label{fig:example_alg1}
\end{figure*}

\paragraph{\bf Example of Transaction Processing Using Algorithm \ref{alg:centralized-scheduler}}

Consider four shards, $S_1$ to $S_4$, each with one pending transaction: $T_1$ to $T_4$ (shown in Figure \ref{fig:example_alg1} (a)). The transactions access accounts on different shards as follows: $T_1$ accesses $S_2$ and $S_3$, $T_2$ accesses $S_2$ and $S_4$, $T_3$ accesses $S_1$ and $S_3$, and $T_4$ accesses $S_3$ and $S_4$. Since each shard holds one account, any transactions that access the same shard will conflict and must be scheduled in different rounds.

Consider shard $S_1$ acts as the leader ($S_\ell$). During Phase~1, each shard sends its transaction information to the $S_\ell$, which constructs a transaction conflict graph $G_\T$ (as shown in Figure~ \ref{fig:example_alg1} (b)). Using a greedy coloring algorithm, the leader assigns colors to the transactions based on conflicts: $T_1$ receives blue, $T_2$ and $T_3$ both receive green, and $T_4$ receives orange (shown in Figure \ref{fig:example_alg1} (c)). The first two rounds are used to share transaction information and to color. Then, the blue transaction $T_1$ executes in rounds $3-6$ (i.e., start at round $3$ and completed by $6$ as each color takes $4$ rounds), the green transactions ($T_2$, $T_3$) in rounds $7-10$, and the orange transaction ($T_4$) in rounds $11-14$.
In Phase 2, each home shard splits its transaction into subtransactions and sends them to the destination shards for processing based on the assigned rounds and colors. For instance, $S_1$ splits $T_1$ into $T_{1,2}$ and $T_{1,3}$, sending them to $S_2$ and $S_3$, respectively (shown in Figure \ref{fig:example_alg1} (d)). Transactions are executed in their designated rounds, with all transactions completed by round 14.

\ra{In the following, we analyze the time required to process transactions in Algorithm \ref{alg:centralized-scheduler}, focusing on the period after the leader shard has determined the schedule for all transactions. More specifically, we omit the cost of Phase 1 in this analysis. If the shard is in a clique graph with uniform distances between shards, this cost remains constant. However, if the distance between a home shard and the leader shard is $d$, an additional cost proportional to $d$ is incurred.}

\subsection{Analysis for General Graph}
Suppose we run Algorithm \ref{alg:centralized-scheduler} for general graph $G_s$.
We have the following results.

\begin{lemma}[Lower Bound]
\label{lemma:lower-bound-general-graph}
    For graph $G_s$, the processing time of any schedule for $\T$ in a graph $G_\T$ is at least $l$, where $l$ is the maximum number of transactions that access any object $o_j\in \cO$.
\end{lemma}
\begin{IEEEproof}
    Let $\T_{o_j}$ denote the set of transactions that use object $o_j \in \cO$. Let $l_{o_j} = |\T_{o_j}|$, and $l=\max_j l_{o_j}$ be the maximum 
    number of transactions that access any object in $\cO$. Then, the length of the schedule is at least $l$ because an object has to be accessed once at a time by $l$ transactions. Consequently, the processing time of any schedule for $\T$ is at least $l$.
\end{IEEEproof}

\begin{lemma}[Upper Bound]
\label{lemma:upper-bound-general-graph}
    For the general graph, the time to process the transactions $\T$ is at most $O(kld)$, where $l$ is the maximum number of transactions that access an object $o_j\in \cO$, $k$ is the maximum number of objects accessed by each transaction, and $d$ is the maximum distance between a transaction and its accessing objects.
\end{lemma}

\begin{IEEEproof}
The
maximum degree in the conflict graph $G_\T$ is at most $kl$, 
since each transaction accesses at most $k$ objects and each object is accessed by at most $l$ transactions.
Therefore, the conflict graph  $G_\T$ can be greedily colored 
with at most $\xi \leq kl+1$
colors. 

The distance between a transaction and its accessing object is at most $d$ far away.
Since Algorithm \ref{alg:centralized-scheduler} has at most $4$ rounds of interactions between a transaction and its accessing objects in Phase 2, each color corresponds to $4d$ communication rounds.
Thus, it takes at most $(kl+1)4d = O(kld)$ rounds to confirm and commit the transactions.

\end{IEEEproof}




\begin{theorem}
\label{non-uniform-approximation-alg1}
     For the general graph $G_s$, Algorithm \ref{alg:centralized-scheduler} gives an $O(kd)$ approximation to the optimal schedule. 
     
\end{theorem}
\begin{IEEEproof}
From Lemma \ref{lemma:upper-bound-general-graph}, the time to process all the transactions is at most $O(kld)$. Moreover, from Lemma \ref{lemma:lower-bound-general-graph}, $l$ is the lower bound, which gives us $O(kd)$ approximation to the optimal schedule.

     

\end{IEEEproof}




Let us consider shards are connected in a clique (complete) graph  $G_s$ with $s$ shards where every shard is connected to every other shard with a unit distance (thus, $d = 1$). Then, from Theorem \ref{non-uniform-approximation-alg1}, Algorithm \ref{alg:centralized-scheduler} gives $O(k)$ approximation to the optimal schedule.





Hence, from Theorem \ref{non-uniform-approximation-alg1} we obtain the following corollaries.

\begin{corollary}
 For a clique graph with unit distance per edge (i.e. $d=1$), Algorithm \ref{alg:centralized-scheduler} gives an $O(k)$ approximation to the optimal schedule.
\end{corollary}

\new{
Let $G_s$ be a Hypercube graph \cite{leighton2014introduction}. Then, there is a connecting path between each pair of shards with length $\log s$; hence $D = O(\log s)$, and $d \leq D = O(\log s)$.
Therefore, Theorem \ref{non-uniform-approximation-alg1} implies that 
there is an execution schedule which is a $O(k \log s)$ approximation of the optimal schedule.
A similar property holds for the Butterfly, and $g$-dimensional grids graph~\cite{chan1989embedding}, where $g=O(\log s)$.
Thus, we have:



\begin{corollary}
 Algorithm \ref{alg:centralized-scheduler} gives an $O(k \log s)$ approximation for Hypercube, Butterfly, and $g$-dimensional grids graph, where $g=O(\log s)$.
\end{corollary}}

 \section{Centralized Scheduler with Buckets}
 \label{sec:centralized-scheduler-with-bucket}

In the Centralized scheduler (Algorithm \ref{alg:centralized-scheduler}), a single shard (the leader shard) colors the transactions and determines the schedule. We propose an alternative algorithm that uses buckets to organize transactions based on the distances between the transaction and its accessed objects. This approach offers improved bounds for special cases when the transaction accesses $k$ objects randomly, as well as for the Line graph.

Here, we distribute the coloring load in different shards, i.e., each bucket has its own leader shard, which is known to all the shards and determines the schedule to process the transactions of that bucket.
We divide the transactions into buckets $B_0, B_1, \ldots$,
according to the maximum distance between the home shard of the transaction and its accessing objects (i.e., destination shards).
In bucket $B_i$, we add transactions whose maximum distance between 
home shard and its destination shard is in range $[2^i, 2^{i+1})$.
We compute a schedule for each bucket.  We also prioritize the processing of the lower index bucket transactions over the higher index bucket so that we can get the locality benefits and reduce transaction latency.
That is, first, we calculate a schedule and process transactions  of $B_0$, then for $B_1$, then for $B_2$,
and so on.

The pseudocode of the centralized scheduler with the bucketing approach is presented in Algorithm \ref{alg:centralized-with-bucket}. We assume that shard graph $G_s$, a set of transactions $\T$ that need to be processed, are given as input to the algorithm. Then, the scheduler provides the schedule and processes all the transactions $\T$. We assume that each bucket $B_i$ has its own leader shard $S_{\ell_i}$, which is the home shard of some transaction in $B_i$ and is known to all shards in $G_s$. First, each transaction determines its bucket, and then we run Algorithm \ref{alg:centralized-scheduler} for each bucket by providing the shard graph $G_s$, set of transactions of each bucket $B_i$ and leader of that bucket as parameters. After completing the transaction processing of bucket $B_i$, the leader of $B_i$ informs the leader of $B_{i+1}$ to run the Algorithm \ref{alg:centralized-scheduler} and process their transactions.


\begin{algorithm}[t]

\small
\caption{{\sc SchedulerWithBuckets}}
\label{alg:centralized-with-bucket}
\SetKwInOut{Input}{Input}\SetKwInOut{Output}{Output}
\Input{Shard graph $G_s$; set of txns $\T$; 
}
\BlankLine

$B_i$ is the set of txns whose distance between the home shard and the accessing objects (destination shards) is in the range $[2^i,2^{i+1})$; $i\in \{0,\dots, \lfloor \log D \rfloor\}$ where $D$ is the diameter of shards graph $G_s$\;
Each Bucket $B_i$ has its own leader shard $S_{\ell_i}$ which is the home shard of some transaction in $B_i$, and it is known to all shards in $G_s$\;


\BlankLine
\BlankLine
\tcc{Run Algorithm \ref{alg:centralized-scheduler} for each bucket $B_i$}
 \SetKwBlock{DoInRound}{\normalfont {{\bf for} {\em Bucket $B_i \leftarrow B_0$} {\bf to} {\em Bucket $B_{ \lfloor \log D \rfloor}$} {\bf do}}}{}
\DoInRound{
Run {\sc CentralizedScheduler}($G_s$, $B_i$, $S_{\ell_i}$)\;
}

\end{algorithm}

\subsection{Analysis for General Graph with Random Object Accesses}
Suppose each shard contains one transaction that needs to be processed, and each transaction accesses at most $k$ objects picked randomly from different shards. 
We consider the case where there is one object per shard,
and each shard is the home of one transaction, that is, 
there are $s$ objects and transactions.
We use the following version of Chernoff bound in our analysis.

\begin{lemma}[Chernoff bound]
	\label{lemma:chernoff}
	Let $X_1,X_2,\cdots,X_{m}$ be independent Poisson trials 
	such that, for $1\leq i\leq m$, 
	$X_i \in \{0,1\}$.
	Then, for $X=\sum_{i=1}^{m}X_i$, $\mu=E[X]$,
	and any $\delta > 1$,
	$\Pr[X \geq (1+\delta) \mu] \leq e^{-\frac{\delta \mu} 2}$,
	and any $0 \leq \delta \leq 1$,
 $\Pr[X \geq (1+\delta) \mu] \leq e^{-\frac{\delta^2 \mu} 3}$
    and $\Pr[X \leq (1-\delta) \mu] \leq e^{-\frac{\delta^2 \mu} 2}$.
\end{lemma}

\begin{lemma}
\label{lemma:maximum-degree-of-graph}
    If a transaction chooses $k$ objects randomly and uniformly (out of $s$ shards), then the maximum number of transactions accessing any object is $l \leq 2k + 4\ln s$, with probability at least $1 - 1/s$.
\end{lemma}

\begin{IEEEproof}
    Consider an object $o_z \in \cO$.
    Let $X^{(z)}_i$ be the random variable such that $X^{(z)}_i = 1$ if transaction $T_i$ accesses object $o_z$. 
    Let $Y_z = \sum_{i=1}^{s} X^{(z)}_i$, and let $\mu=E[Y_z]$ be the expected number of transactions that access object $o_z$. 
    

    Using the union bound, the probability that there exists an object $o_z$ such that  $Y_z$ exceeds a certain value $c$ is at most the sum of the probabilities that each $Y_z$ exceeds $c$.
    $$
    \mathrm{Pr}(l \geq c) = \mathrm{Pr}(\text{exists $o_z$  s.t. } Y_z \geq c)  \leq \sum_{z=1}^{s} \mathrm{Pr}(Y_z \geq c) \ .
    $$

    
    Since each transaction accesses $k$ objects, and each shard has a transaction, the expected number (average) of transactions accessing each object is $\frac{s \cdot k}{s}= k$.

    From Lemma \ref{lemma:chernoff} (Chernoff bound) for a random variable
    $Y_z$ and any $\delta>1$ with expected value $\mu = k$ we have:
    $$
         \mathrm{Pr}(Y_z \geq (1 + \delta)\mu) \leq e^{-\frac{\delta \mu}{2}} \ .
    $$

    We want to find the probability that the number of transactions accessing this object is at least $c$, i.e., 
    $\mathrm{Pr}(Y_z \geq c)$. To apply the Chernoff bound, we set  $c=(1+\delta)k$, which gives us $\delta = \frac{c-k}{k}$.
    We substitute 
    these into the Chernoff bound: 
    $$\mathrm{Pr}(Y_z \geq c) \leq e^{-\frac{\frac{(c-k)}{k}}{2}k}= e^{-\frac{(c-k)}{2}} \ .$$
    For the union bound, we sum these probabilities over all objects:
$$
\mathrm{Pr}(l \geq c) \leq \sum_{i=1}^{s} \mathrm{Pr}(Y_z \geq c)
= s \cdot e^{-\frac{(c-k)}{2}} \ .
$$

We want this probability to be less than or equal to $\frac{1}{s}$, so
$
s \cdot e^{-\frac{(c-k)}{2}} \leq \frac{1}{s}.
$.
Solving for $c$, we get:
$
c \geq k +  4\ln s
$.
We assume $c=(1+\delta)k$, where $\delta>1$. Thus, 
we can take 
$
c \geq 2k +4\ln s
$.
Therefore, $l \leq 2k + 4\ln s$ with probability at least $1 - \frac{1}{s}$.


\end{IEEEproof}




    

\begin{lemma}[Lower Bound]
\label{lemma:lower-bound-random-object}
    The time to process the transactions in bucket $B_i$ is at least
    $2^i$.
\end{lemma}

\begin{IEEEproof}
This follows immediately from the definition of bucket $B_i$,
since there is a transaction in $B_i$ that has distance at least $2^i$ to one of its accessing objects.
\end{IEEEproof}

 Suppose $|B_i|$ is the total number of transactions in the bucket $B_i$. Let $B_i(o_x) \subseteq B_i$ denote the set of transactions in $B_i$ that use object $o_x\in \cO$. Let $l_i=\max_{o_x \in \cO} |B_i(o_x)|$ be the maximum number of transactions in $B_i$ that used any object in $\cO$.

\begin{lemma}[Upper Bound]
\label{lemma:upper-bound-random-object}
    If each transaction accesses at most $k$ objects randomly from different shards, then the time to process the transactions in any bucket $B_i$ is at most 
    $2^{i+3}(k(2k + 4\ln s) +1)$, with probability at least $1 - 1/s$.
\end{lemma}
\begin{IEEEproof}
     From Lemma \ref{lemma:maximum-degree-of-graph}, if a transaction picks to access $k$ objects randomly, then an object is used by at most $2k + 4\ln s$ transactions (with probability at least $1 - 1/s$). Thus, we can write $l_i \leq  2k + 4\ln s$.  

     Let $G_{B_i}$ denote the transaction conflict graph for the transactions in bucket $B_i$. 
     Since all the transactions in the bucket $B_i$ access at most $k$ objects and each object is used by at most $l_i$ transactions, the maximum degree in the transaction graph of bucket $B_i$ is at most $kl_i$. 
     Therefore, the conflict graph $G_{B_i}$ 
     can be greedily colored with $\xi \leq kl_i+1$ colors.

     For a transaction in $B_i$, the distance to its accessing object is at most $2^{i+1}-1 <2^{i+1}$. Moreover, the execution of Algorithm~\ref{alg:centralized-scheduler} has at most $4$ rounds of interactions between the transaction and its accessing objects for the confirmation and commit. Therefore, each color corresponds to at most $4\cdot 2^{i+1}$ rounds. Thus, it takes at most 
     \begin{equation*}
     4\cdot2^{i+1} \xi \leq 4\cdot2^{i+1} (kl_i+1) 
     \leq 2^{i+3}(k(2k + 4\ln s) +1)
     \end{equation*}
     rounds to confirm and commit transactions.
\end{IEEEproof}
     

    




\begin{theorem}
    For the General graph, if each transaction accesses $k$ objects randomly from different shards, then Algorithm \ref{alg:centralized-with-bucket} gives an $O(k \log D \cdot (k + \log s))$ approximation to the optimal schedule with probability at least $1-1/s$.
    \label{approximation-for-general-graph}
\end{theorem}

\begin{IEEEproof}
  Let $B_m$ be the bucket that takes the most time to process all its transactions.
  Thus, when we consider all the $\lfloor \log D \rfloor+1$ buckets,
 from Lemma \ref{lemma:upper-bound-random-object}, 
 the total number of rounds needed to process all transactions is at most:
\begin{eqnarray*}
\label{eqn:combined-bucket-cost}
  & & \sum_{i=0}^{\lfloor \log D \rfloor+1} 2^{i+3}(k(2k + 4\ln s) +1)\\
  & & \leq 2^{m+3}(k(2k + 4\ln s) +1) (\log D+2)\ .
\end{eqnarray*}

From Lemma \ref{lemma:lower-bound-random-object}, $2^m$ is a lower bound on the cost of processing the transactions in bucket $B_m$.
Therefore,
the algorithm approximates the optimal cost within a factor
$$ \frac{2^{m+3}(k(2k + 4\ln s) +1) (\log D+2)} {2^m} = O((k^2 + \log s)\log D) \ .$$
\end{IEEEproof}


\subsection{Analysis for Line Graph with Buckets}


We consider graph $G_s = (V,E)$ which is a line graph where shards are connected in the sequence of $s$ shards, $S_1, S_2, \cdots S_s \in V$,
where any two consecutive shards $S_i,S_{i+1}$ are connected with an edge of weight $1$, for $1\leq i \leq s-1$. 
Generally, for any pair $S_i, S_j$, where $i < j$,
there is an edge of weight $j-i$, where $1 \leq i,j \leq s$. 
We consider the orientation of the shards in the line to be $S_1$ the leftmost and $S_s$ the rightmost. We continue with the analysis where we use the buckets as defined earlier, and we assume that each shard has one transaction to be processed.


\begin{lemma}[Lower Bound]
\label{lemma:line-lower-bound}
    For the Line graph, the time to process the transactions in a non-empty bucket $B_i$ is at least $2^i + \frac{l_i^2}{8}$.
\end{lemma}
\begin{IEEEproof}
    Let $o_j$ be the object for which the parameter $l_i$ is maximized.
    Let $B_i' = B_i(o_j)$,
    with $l_i=|B_i'|$.  
     We can decompose the bucket $B_i'$ into partial buckets $\bar{B}'_{0,i}, \bar{B}'_{1,i}, \cdots, \bar{B}'_{z,i}$ where partial bucket $\bar{B}'_{z,i}$ contains the transactions which access object $o_j$ from distance $[2^z, 2^{z+1})$.
     
     Without loss of generality, suppose the object $o_j$ is in the middle of the line (assuming $s$ is odd). Since each shard hosts a single transaction,
     we have
     $|\bar B_{z,i}'| \leq 2^{z+1} - 2^z = 2^z$.
     The number of transactions up to $z^* = \lceil (\log l_i)-3 \rceil$ (logarithm is base 2) is
     \[
        \sum_{z=0}^{\lceil (\log l_i)-3 \rceil} |\bar B_{z,i}'| \leq \sum_{z=0}^{(\log l_i)-2}2^{z}\leq 2 \cdot 2^{(\log l_i)-2} \leq \frac{l_i} {2} \ .
     \]
     
     As the transactions are in line, the remaining $l_i/2$ transactions must be in the bucket index $\bar{B}'_{>z^*,i}$. 
     Bucket $\bar{B}'_{z^*+1,i}$ has transactions at distance at least $2^{z^*+1} \geq 2^{\lceil (\log l_i)-3 \rceil + 1} \geq 2^{(\log l_i) - 2} \geq l_i / 4$.    
     This implies that these $l_i/2$ transactions are at least $l_i/4$ distance from object $o_j$.
     
     Thus, $l_i/2\cdot l_i/4=\frac{l_i^2}{8}$ is a lower bound on the execution time because each of these $l_i/2$ transactions need to be serialized (because the conflict on each other on $o_j$), and they also need to travel at least $l_i/4$ distance to reach object $o_j$.
    
    Moreover, there is at least one transaction in bucket $B_i$, which accesses the object at a distance of at least $2^i$; that is why they are in bucket $B_i$. Consequently, the total time to process all the transactions of bucket $B_i$ is at least $2^i+\frac{l_i^2}{8}$.
\end{IEEEproof}

The distance between a transaction in $B_i$ 
and its accessing objects is at most $2^{i+1}-1 < 2^{i+1}$.
Thus, from Lemma \ref{lemma:upper-bound-general-graph},
since $d \leq 2^{i+1}$,
we get the following result.

\begin{lemma}[Upper Bound]
\label{lemma:line-upper-bound}
      For the Line graph, the time to process the transactions in bucket $B_i$ is at most $O(2^ikl_i)$.
\end{lemma}



\begin{theorem}[Approximation]
For the Line shard graph where each transaction accesses at most $k$ objects, Algorithm~\ref{alg:centralized-with-bucket} provides an $O(k\sqrt{d} \log D)$ approximation to the optimal schedule.
\end{theorem}
\begin{IEEEproof}
Let $B_m$, $0 \leq m \leq \lfloor \log D \rfloor$, be the bucket 
that maximizes the term $2^mkl_m$.
Since we have $\lfloor \log D \rfloor + 1= O(\log D)$ buckets,
    from Lemma \ref{lemma:line-upper-bound} the time to process the transactions in all the buckets is at most $O(2^m kl_m \log D)$. 
    From Lemma \ref{lemma:line-lower-bound}, the time to process the transactions of bucket $B_m$ is at least $2^m+l_m^2/8 = \Omega(2^m+l_m^2)$.
    Therefore, we obtain an 
    $O \left(\frac{2^mk l_m \log D}{2^m + l_m^2}\right )$ approximation factor for time.
    We analyze two cases:
    
    (i) $l_m< 2^{m/2}$:
    \begin{multline}
    \frac{2^{m}k l_m \log D}{2^m + l_m^2} 
    \leq \frac{2^{m}k 2^{m/2} \log D}{2^m} 
    = k \sqrt{2^m}\log D\nonumber \ .
    \end{multline}
    By the definitions of $B_m$ and $d$, $2^{m+1} \leq d$.
    Thus, we get $O(k\sqrt{d}\log D)$ approximation factor.
    
    (ii) $l_m \geq 2^{m/2}$:
    \begin{multline}
        \frac{2^{m}k l_m \log D}{2^m + l_m^2} \leq \frac{2^{m}k l_m\log D}{ l_m^2} = \frac{2^{m}k \log D}{ l_m}\\ \leq  k \sqrt{2^m}\log D  = O(k\sqrt{d}\log D) \nonumber \ .
    \end{multline}
    Combining cases (i) and (ii) 
    we get that Algorithm \ref{alg:centralized-with-bucket} gives an $O(k\sqrt{d} \log D)$ approximation to the optimal schedule.
\end{IEEEproof}


\section{Distributed Scheduler (DS)}
\label{sec:fully-distributed}

The scheduling algorithms we presented earlier use a central authority in a shard that has knowledge about all current transactions and the maximum degree of the transaction graph. However, such a central authority might not exist in blockchain sharding since each transaction is generated in a distributed manner in any shard. Here, we discuss a fully distributed scheduling approach using a clustering technique that allows the transaction schedule to be computed in a decentralized manner without requiring a single central authority. Each cluster has its own leader and within each cluster the distributed approach invokes centralized scheduling Algorithm \ref{alg:centralized-scheduler}.




\subsection{Shard Clustering}
\label{shard-clustering}
We define the $z$-neighborhood of shard $S_i$ 
as the set of shards within a distance of at most $z$ from $S_i$. Moreover, the 0-neighborhood of shard $S_i$ is just the $S_i$ itself.

We consider that our distributed scheduling algorithm uses a hierarchical decomposition of $G_s$
which is known to all the shards and calculated before the algorithm starts.
This shard clustering (graph decomposition) is based on the clustering techniques in \cite{gupta2006oblivious} and which were later used in \cite{sharma2014distributed,busch2022dynamic, adhikari2024spaastable}.
We divide the shard graph $G_s$ into the hierarchy of clusters with $H_1 = \lceil \log D \rceil +1$ layers (logarithms are base 2). A layer is a set of clusters, where a cluster is a set of shards. 
Layer $q$, where $0\leq q < H_1$, is a sparse cover of $G_s$
such that:
\begin{itemize}
    \item Every cluster of layer $q$ has diameter of at most $O(2^q\log s)$.
    \item Every shard participates in no more than $O(\log s)$ different clusters at layer $q$.
    \item For each shard $S_i$ there exists a cluster at layer $q$ which contains the $(2^q-1)$-neighborhood of $S_i$ within that cluster.
\end{itemize}
 

For each layer $q$,
the sparse cover construction in \cite{gupta2006oblivious}
is actually obtained as a collection of $H_2 = O(\log s)$ partitions of $G_s$. 
These $H_2$ partitions are ordered as sub-layers of layer $q$ labeled from $0$ to $H_2-1$.
A shard might participate in all $H_2$ sub-layers, 
and at least one of these $H_2$ clusters at layer $q$ contains the whole $2^q-1$ neighborhood of $S_i$.

In each cluster at layer $q$, a leader shard is specifically designated such that the leader’s $(2^q-1)$-neighborhood is in that cluster.
To express layers and sub-layers of a cluster, we define the concept of {\em height} as a tuple $h= (h_1, h_2)$, where $h_1$ denotes the layer and $h_2$ denotes the sublayer of the cluster. Similar to \cite{sharma2014distributed, busch2022dynamic,  adhikari2024spaastable}, heights are ordered lexicographically.

The {\em home cluster} for each transaction $T_i$ is defined as follows: suppose $S_i$ is the home shard of $T_i$,
and $z$ is the maximum distance from $S_i$ 
to the destination shards that will be accessed by $T_i$;
the home cluster of $T_i$ is the lowest-layer (and lowest sub-layer) cluster 
in the hierarchy that contains the $z$-neighborhood of $S_i$.
Each home cluster has a dedicated leader shard, which will handle all the transactions that have their home shard in that cluster (i.e., transaction information will be sent from the home shard to the cluster leader shard to determine the schedule).

\subsection{Distributed Scheduling Algorithm}
In the distributed scheduler Algorithm \ref{alg:fully-distributed-scheduler}, each home shard determines the home cluster for its transaction based on the distance between the home shard of the transaction and its accessing objects (destination shards) and sends that transaction information to the leader of the respective home cluster.
Each cluster $C$ belongs to some layer $q$ and sublayer $r$ (height $(q,r)$),
where $0 \leq q < H_1$ and $0 \leq r < H_2$.
Suppose a transaction $T_i$ has home cluster $C$;
we also say that $T_i$ is at height $(q,r)$. 
The leader shard of $C$, denoted $S_{\ell}(C)$, 
assigns an integer {\em color} to each transaction $T_i$.
We define the concept of the {\em height} for $T_i$ represented by a tuple $(q, r, \text{color})$.

Using the lexicographic order of heights, we implement a priority scheme for the schedule. The priority for committing transactions is determined based on this order, giving precedence to transactions generated in lower cluster layers (and within a layer priority is given to lower sublayers). The reason behind this priority scheme is to take the benefit of the locality to improve the transaction latency. Note that a home shard may have transactions at various heights.
Those transactions will be processed at their respective heights.

For each level $(q,r)$, each cluster $C$ invokes Algorithm~\ref{alg:centralized-scheduler} parallelly to schedule and process the transactions. While invoking Algorithm \ref{alg:centralized-scheduler} for each cluster $C$ we set the parameters as $G_s[C]$, $\T(C)$, $S_\ell(C)$, where $G_s[C]$ is the cluster graph induced by $C$ (i.e. containing the shards in $C$), $\T(C)$ is the set of transactions whose home cluster is $C$, and $S_\ell(C)$ is the leader shard of the cluster $C$.

\begin{algorithm}[t]
\small
\caption{ {\sc DistributedScheduler}}
 \label{alg:fully-distributed-scheduler}
\SetKwInOut{Input}{Input}\SetKwInOut{Output}{Output}
\Input{Shard graph $G_s$; set of txns $\T$; 
}

Assume a hierarchical cluster decomposition of $G_s$, which is known to all the shards in $G_s$\;  

Each cluster $C$ belongs at some height $(q,r)$, at layer $q$ and 
sublayer $r$ of the hierarchical decomposition, where $0 \leq q < H_1$ and $0 \leq r < H_2$; the heights $(q,r)$ are ordered lexicographically\;


    
The home shard of each txn $T_i \in \T$ discovers the position of its destination shards in $G_s$\;

The home shard of each txn $T_i$ picks for home cluster the lowest height $(q,r)$ cluster $C'$, which includes $T_i$ and the destination shards (which are at distance at most $z$) of its subtransactions\;



\BlankLine
\BlankLine

\tcc{Run Algorithm \ref{alg:centralized-scheduler} for each cluster}
\SetKwBlock{DoParallel}{\normalfont{ {\bf for each}  {\em cluster layer $q \gets 0 $ to $H_1$ and sublayer $r \gets 0$ to $H_2$ }{\bf do}}}{end}

    \DoParallel{
     \SetKwBlock{DoInRound}{\normalfont {{\bf for each} {\em cluster $C$ in level $(q,r)$} {\bf in parallel do}}}{}
\DoInRound{
$\T(C) \leftarrow$ set of transactions with home cluster $C$\;
$S_\ell(C)$ $\leftarrow$ leader shard of cluster $C$\;
$G_s[C] \leftarrow$ cluster graph induced by $C$\;
Run {\sc CentralizedScheduler}($G_s[C]$, $\T(C)$, $S_\ell(C)$)\;
}
    }


                
                
                

\end{algorithm}

\subsection{Invocation of Centralized Scheduler}
\label{simulation-of-alg2-from-alg1}

The distributed scheduler (Algorithm \ref{alg:fully-distributed-scheduler})
invokes the centralized scheduler (Algorithm \ref{alg:centralized-scheduler}) in each home cluster. By invoking Algorithm \ref{alg:centralized-scheduler}, the leader shard $S_\ell(C)$
in a cluster $C$ knows the transactions assigned to it and schedules them in a conflict-free manner using vertex coloring of $G_s[C]$. 


We can also use (Algorithm \ref{alg:centralized-with-bucket}), instead of Algorithm \ref{alg:centralized-scheduler},
to further improve the bounds for the Line graph and for special cases in the general graph when the \( k \) objects are selected randomly.
Thus, for each cluster \( C \), we can run Algorithm \ref{alg:centralized-with-bucket}, which will ultimately invoke Algorithm \ref{alg:centralized-scheduler}. 

Next, we show that due to the layer and sublayer structure of the cluster hierarchy, we need to pay an \( O(\log d \cdot \log s) \) factor cost over the approximation obtained in the centralized approach, which is proven in the following theorem.

\begin{theorem}
\label{simulation-proof}
    The distributed scheduler (Algorithm \ref{alg:fully-distributed-scheduler}) has a competitive ratio of $O(\mathcal{A}_{\mathcal{CS}} \cdot \log d \cdot \log s)$, where $\mathcal{A}_{\mathcal{CS}}$ is the approximation ratio of the centralized scheduling algorithm ($\mathcal{CS}$).
\end{theorem}
\begin{IEEEproof}
    Let $C$ be a cluster that takes the maximum time $\tau$  to process transactions compared to any other cluster in the hierarchy. Then, we can write
    $$\tau \leq \mathcal{A}_{\mathcal{CS}} \cdot \tau^*$$ where
    $\tau^*$ is the optimal scheduling time for $C$.

Now, we consider all the $H_1$ layers and $H_2$ sublayers within each layer.
The algorithm will only use transactions up to a layer $q' = O(\log d)$,
since the clusters at that layer cover all the transactions and the respective
objects that they access.
Thus, the total rounds required by Algorithm \ref{alg:fully-distributed-scheduler} to process all the transactions from all cluster level $(q,r)$ 
is at most
\begin{equation}
\label{eqn-combined-layer-sublayer}
 \tau_{total} \leq \sum_{q=0}^{q'} \sum_{r=0}^{H_2-1} \tau \leq \sum_{q=0}^{q'} \sum_{r=0}^{H_2-1}\mathcal{A}_{\mathcal{CS}} \cdot \tau^* 
 = \mathcal{A}_{\mathcal{CS}} \cdot \tau^* \cdot c \log d \cdot \log s\nonumber 
\end{equation}
for some constant $c > 0$.
Since $\tau^*$ is the lower bound for any scheduler, we can calculate the approximation ratio as:

$$\frac{\tau_{total}}{\tau^*} \leq \frac{\mathcal{A}_{\mathcal{CS}} \cdot \tau^* \cdot c \log d \log s}{ \tau^*} = O(\mathcal{A}_{\mathcal{CS}} \log d \cdot \log s) \ .$$

\end{IEEEproof}

Since $\mathcal{A}_{\mathcal{CS}}$ is the approximation ratio of the centralized scheduling algorithm, from Theorem \ref{simulation-proof}, we conclude that we need to pay an extra $O(\log d \cdot \log s)$ rounds penalty for the distributed scheduler (Algorithm \ref{alg:fully-distributed-scheduler}) due to hierarchical clustering. Hence, we have the following results:

\begin{itemize}
    \item  For a General graph, Algorithm \ref{alg:fully-distributed-scheduler} provides an $O(kd \cdot \log d \cdot \log s)$ approximation to the optimal schedule.
    \new{
    \item For Hypercube, Butterfly, and $g$-dimensional grids graphs, Algorithm \ref{alg:fully-distributed-scheduler} provides an $O(k \cdot \log d \cdot \log^2 s)$ approximation to optimal schedule, where $g=O(\log s)$.}
   
    \item  For a General graph, If the transaction accesses  $k$ objects randomly, then Algorithm~\ref{alg:fully-distributed-scheduler} gives an $O(k \log D \cdot (k + \log s) \log d \cdot \log s)$ approximation to the optimal schedule with probability at least $1-1/s$.
     \item  For a Line graph, Algorithm \ref{alg:fully-distributed-scheduler} gives an $O(k \sqrt{d} \log D \cdot \log d \cdot \log s)$ approximation to the optimal schedule.

\end{itemize}

We would like to note that the distributed scheduler needs to pay an additional cost of $O(\log d \cdot \log s)$ compared to the centralized scheduler, but it offers significant benefits. Unlike the centralized scheduler, which relies on a single shard to schedule all transactions, the distributed scheduler assigns this responsibility across multiple shards and determines the schedule in a distributed manner. This distribution mitigates performance bottlenecks and load concentration on one shard. Additionally, by prioritizing transaction processing in lower layers, the distributed scheduler takes advantage of locality to improve the transaction latency.

    





\section{Simulation Results}
\label{sec:simulation-results}
In this section, we provide our simulation results, conducted on a
MacBook Pro with an Apple M1 chip featuring a 10-core CPU, a
16-core GPU, and 32 GB of RAM. The simulation was implemented
using Python programming language with relevant libraries and
resources. We considered the following specific parameters for
simulation: the number of shards, $s$, varies from $16$ to $128$ to observe the effects of increasing the number of shards, with each shard holding one account. The maximum number of shards accessed by each transaction, $k$, was varied with values of $2$, $4$, and $8$. Moreover, we ran our simulation instance ten times and plotted the average results. \new{As our proposed algorithms are lock-free algorithms, we compare our result with the two-phase locking (2PL) approach which is used by Byshard\cite{Byshard}, where accounts are locked for concurrency control.}

Initially, we generated unique accounts (objects) and randomly assigned them to different shards, with each shard holding a unique account (object) that needs to be modified in an atomic and consistent manner. We assume that IoT, edge, or mobile devices generate transactions that attempt to modify the objects stored in the shards. Additionally, we assume that each shard has one transaction that needs to be processed and appended to the blockchain. To simulate this scenario, we generate $s$ transactions, each accessing (attempting to update/modify/store) at most $k$ accounts (objects). In other words, each transaction accesses at most $k$ shards since each shard contains one account. Our simulation runs in rounds until all transactions are processed. We measure the average throughput, average transaction latency, \new{and average number of message exchanges} with respect to the number of shards. Here in this paper, we measure time in rounds. Thus, here, we define throughput based on rounds. More specifically, if we have $\T$ transactions and the number of required rounds to process $\T$ transactions is $y$, then throughput is $\T/y$. So, we say throughput (in transactions per round) and latency (in rounds).

\subsection{Simulation Results of Algorithm \ref{alg:centralized-scheduler}}
\begin{figure*}[!ht]
  
  \centering
  \includegraphics[width=0.9\textwidth]{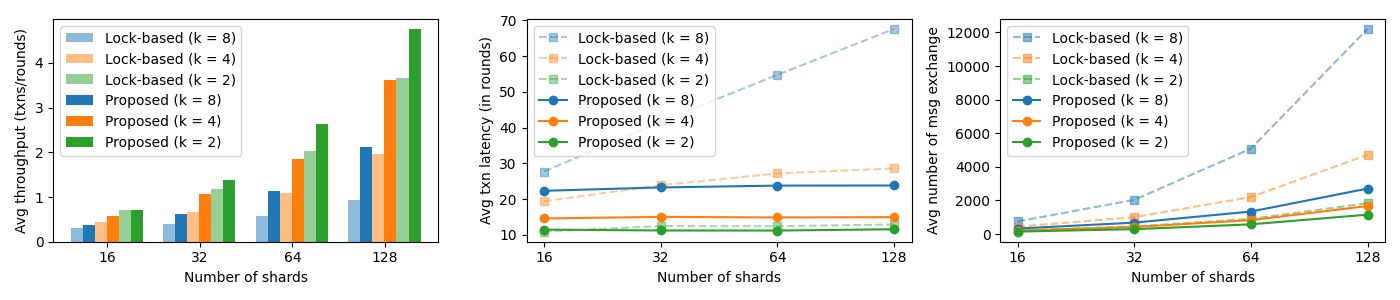}
  \caption{Simulation results of Algorithm \ref{alg:centralized-scheduler}.  The left bar chart shows the average transaction throughput (in transactions per round) versus the number of shards. The middle line graph shows the average transaction latency (in rounds) versus the number of shards, and the right line graph shows the average number of message exchanges versus the number of shards.}
  \label{fig:alg1_throughput_vs_number_of_shards}
\end{figure*}

In this simulation, we configured the shards so that each shard is connected to every other shard (i.e., the shards form a clique graph) with a unit distance. This means that any shard can send or receive transaction information to or from any other shard within one round. Figure \ref{fig:alg1_throughput_vs_number_of_shards} shows the simulation results of Algorithm \ref{alg:centralized-scheduler} \new{compared to the Lock-based approach (Byshard \cite{Byshard})}. The left bar chart displays the average transaction throughput (in transactions per round) versus the number of shards, the middle line graph represents the average transaction latency (in rounds) against the number of shards, \new{and the right line graph shows the average message exchanges with respect to the number of shards}. As expected, as the number of shards increases, the average transaction throughput also increases, with little effect on transaction latency and message exchanges because independent transactions are processed and committed in parallel across the shards. 

\new{It is important to highlight that our proposed scheduling algorithm outperforms the Lock-based protocol (Byshard \cite{Byshard}) in terms of throughput, latency, and the number of message exchanges. While the performance difference is minimal when the number of shards is $16$, and the transaction accesses only two shards (i.e., $k=2$), the Lock-based approach exhibits significantly worse performance as the number of shards ($s$) increases or the number of shards accessed by each transaction ($k$) increases.
For example, when transactions access at most $k=8$ shards and the total number of shards is $s=128$, our algorithm achieves an average throughput greater than $2$, compared to just $1$ for the Lock-based approach, which makes our method twice as effective. Similarly, at $s=128$ and $k=8$, the average latency for our scheduling algorithm is approximately $22$ rounds, whereas the Lock-based approach has an average latency of about $70$ rounds. This demonstrates that our approach provides three times better latency.

Moreover, the Lock-based approach requires a significantly higher number of message exchanges because it needs to handle contention when accounts (objects) accessed by a transaction are locked by other transactions. Additionally, it implements prioritization to avoid deadlocks, which increases communication overhead \cite{Byshard}. For the same configuration ($s=128$ and $k=8$), the Lock-based protocol generates around $12,000$ message exchanges, while our approach requires only $2000$, making our method $6$ times more efficient in terms of communication overhead. These results demonstrate the scalability and efficiency of our proposed scheduling algorithm.}

\subsection{Simulation Results of Algorithm \ref{alg:centralized-with-bucket}}
To simulate the algorithm \ref{alg:centralized-with-bucket}, we arranged the shards in a line, where the distance between any two adjacent shards is 1. We ran the experiment with varying numbers of shards: 16, 32, 64, and 128. For example, when configuring 16 shards ($S_1$ to $S_{16}$) in a line, the distance between two adjacent shards, such as $S_1$ and $S_2$, is set to 1. The distance increases as we move further along the line, so the distance from $S_1$ to $S_3$ is 2, from $S_1$ to $S_4$ is 3, and so on. To organize transactions and their accessed shards, we applied the bucketing approach described in Algorithm \ref{alg:centralized-with-bucket}. Transactions are divided into buckets $B_0, B_1, \ldots$ based on the maximum distance between a transaction's home shard and its destination shards. Specifically, transactions are placed in bucket $B_i$ if the maximum distance falls within the range $[2^i, 2^{i+1})$. We then compute a schedule for each bucket.

This simulation provides two sets of results: one where transactions access shards randomly, and another where transactions access nearby shards. We will discuss each result in the following sections.

\subsubsection{Transaction Access Random Shards}
\label{sec:alg2_txn_access_random_shard}
\begin{figure*}[!ht]
  \centering
\includegraphics[width=0.9\textwidth]{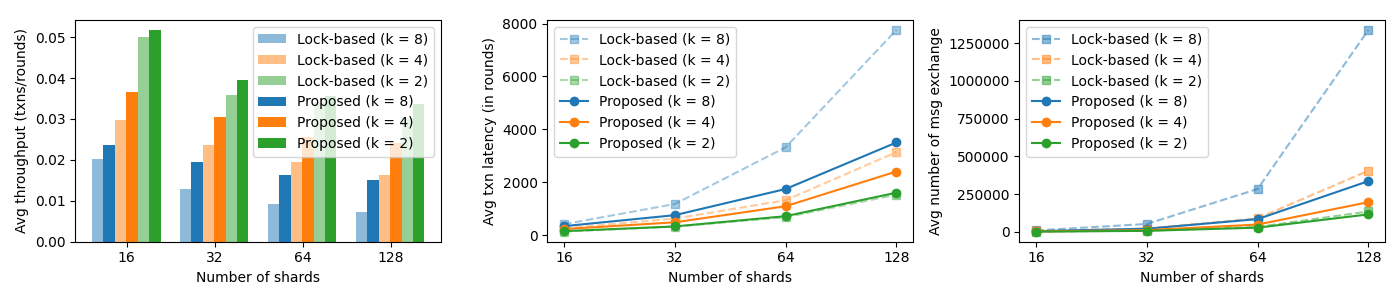}
  \caption{Simulation results of Algorithm \ref{alg:centralized-with-bucket}
for transactions accessing random shards.}
  \label{fig:alg2_txn_access_random_shard}
\end{figure*}

In this simulation, transactions are generated so that each transaction randomly accesses $k$ shards. Figure \ref{fig:alg2_txn_access_random_shard} presents the simulation results, with average transaction throughput on the left, transaction latency on the middle, and an average number of message exchanges on the right.
The results show that increasing the number of shards does not improve throughput, and transaction latency increases significantly. This is because as the number of shards increases, the diameter of the shard line graph also grows, which results in higher communication costs.

For instance, if we have $16$ shards and a transaction accesses $2$ shards—one at the leftmost end and the other at the rightmost end—these shards are $15$ units apart, necessitating back-and-forth communication to validate and commit the transaction (or subtransactions) consistently. When the number of shards increases from $16$ to $32$, this communication distance grows from $15$ to $31$. With $64$ shards, the distance could reach up to $63$ units.

As a result, when transactions access shards randomly, increasing the number of shards actually reduces overall average transaction throughput and increases latency \new{and number of message exchanges. However, we can see from Figure \ref{fig:alg2_txn_access_random_shard} that our proposed scheduling algorithm is better than the lock-based approach for all three performance matrices.}

\begin{figure*}[!ht]
  \centering
\includegraphics[width=0.9\textwidth]{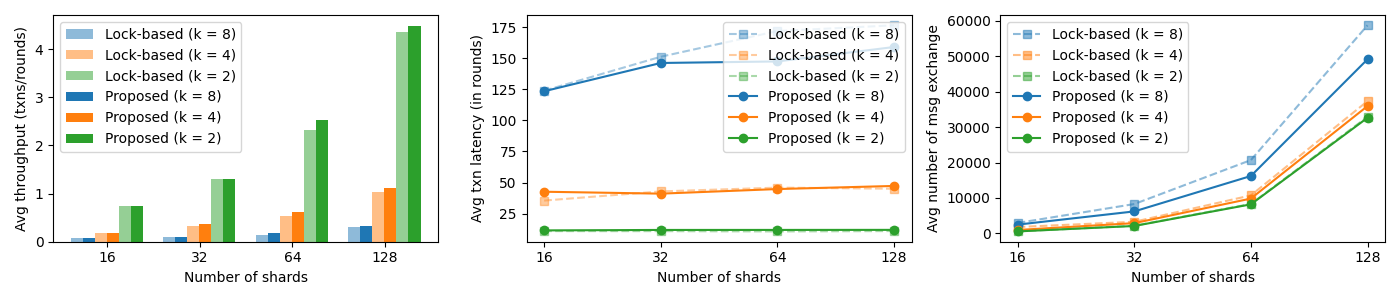}
  \caption{Simulation result of Algorithm \ref{alg:centralized-with-bucket} when transactions access nearby shards.}
  \label{fig:alg2_txn_access_near_by_shard}
\end{figure*}

\begin{figure*}[!ht]
  \centering
\includegraphics[width=0.9\textwidth]{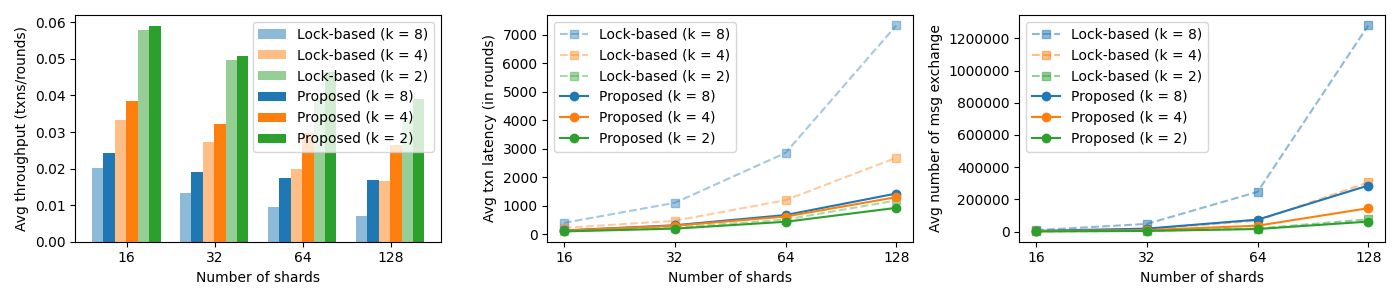}
  \caption{Simulation results of Algorithm \ref{alg:fully-distributed-scheduler}
when transactions accessing random shards.}
  \label{fig:alg3_txn_access_random_shard}
\end{figure*}

\begin{figure*}[!ht]
  \centering
\includegraphics[width=0.9\textwidth]{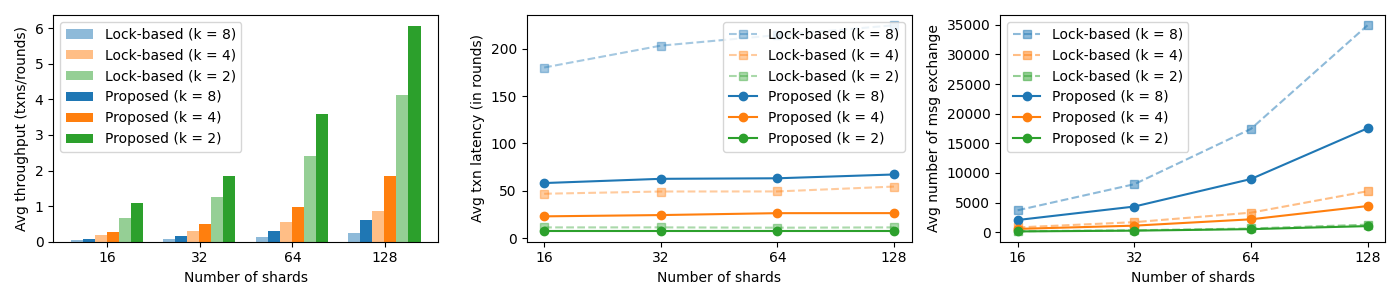}
  \caption{Simulation result of Algorithm \ref{alg:fully-distributed-scheduler} when transactions access nearby shards. }
  \label{fig:alg3_txn_access_near_by_shard}
\end{figure*}

\subsubsection{Transaction Access Nearby Shards}
\label{sec:alg2_txn_access_nearby_shard}

In this scenario, transactions are generated in such a way that each transaction accesses nearby shards, and we analyze the average throughput, latency, \new{and number of message exchanges} as the number of shards increases. For example, if a transaction $T_i$ accesses two shards $S_1$ and $S_2$, these shards are at most one unit distance apart. Similarly, if a transaction accesses four shards $S_1$, $S_2$, $S_3$, and $S_4$, the maximum distance between these shards is at most three units.

More specifically, consider a line graph of 64 shards ranging from $S_1$ to $S_{64}$. If a transaction accesses two shards, one of which is $S_{30}$, the other shard will either be $S_{29}$ or $S_{31}$. This proximity allows for better transaction throughput and lower latency compared to randomly generated transactions. This approach is particularly relevant to mobile and IoT computing, where IoT devices typically generate transactions that access nearby shards\cite{li2021scaling}.

Figure \ref{fig:alg2_txn_access_near_by_shard}
 shows the average throughput, latency, and number of message exchanges where each transaction accesses nearby shards. As the number of shards increases, the throughput also increases; however, if each transaction accesses more shards, the communication cost rises, which leads to reduced throughput improvement. The figure shows that when transactions access only $k=2$ shards, throughput increases significantly. However, when $k=4$ or $k=8$, the increase in throughput is marginal as the number of shards grows. Additionally, our proposed algorithm achieves higher throughput, lower latency, and fewer message exchanges compared to the Lock-based approach for this model.

\subsection{Simulation Results of Algorithm \ref{alg:fully-distributed-scheduler}}
We simulate Algorithm \ref{alg:fully-distributed-scheduler} by clustering the shard as described in Section \ref{shard-clustering}. To organize the shards, we cluster them into layers and sublayers. The sublayers in each layer are constructed by shifting half of the diameter of that layer to the right.
In the lowest layer $l=1$, clusters consist of two shards each. Moving to the next higher layer, $l=2$, clusters now consist of four shards, and this pattern continues. In the highest layer, all shards are part of a single cluster. Each cluster has a designated leader $S_{\text{ldr}}$ responsible for coloring and scheduling all transactions within that cluster. We present the simulation results of this algorithm, where transactions access random shards and nearby shards, respectively.

\subsubsection{Transactions Access Random Shards}
When transactions access random shards, as described in Section \ref{sec:alg2_txn_access_random_shard}, we observe that throughput, transaction latency, and the number of message exchanges do not improve. This lack of improvement is due to transactions potentially accessing shards (or accounts) that are far apart from each other. Figure \ref{fig:alg3_txn_access_random_shard} illustrates the average transaction throughput, latency, and number of message exchanges as the number of shards increases. The simulation results indicate that even with an increasing number of shards, no performance benefits are observed if transactions access two or more shards that are significantly distant from each other. \new{However, lock-based provides worse performance than ours}.

\subsubsection{Transaction Access Nearby Shards}
 The simulation results of Algorithm \ref{alg:fully-distributed-scheduler} where transactions accessing nearby shards is shown in Figure \ref{fig:alg3_txn_access_near_by_shard}. We can observe better throughput, latency, and message exchange from the simulation result. Figure presents the results of the scheduling Algorithm \ref{alg:fully-distributed-scheduler}. The left bar chart shows the average transaction throughput, the middle line graph shows the average transaction latency, and the right line graph illustrates the average number of message exchanges. \new{Simulation results show that our proposed scheduling protocol is better than the lock-based approach.}

\section{Conclusions}
\label{sec:conclusion}

We presented efficient and fast execution time schedules to process batches of transactions for blockchain sharding systems under a synchronous communication model. Our approach includes centralized and distributed schedulers. In the centralized scheduler, we assume each shard has global transaction information, while in the distributed scheduling algorithm, shards do not have global information. We provide optimal approximations for the proposed scheduling algorithm, along with lower and upper bounds. Our results represent the first known attempts to obtain provably fast batch transaction scheduling for blockchain sharding. \new{Additionally, our simulation results demonstrate that the proposed scheduling algorithm outperforms the Lock-based approach in terms of performance metrics such as throughput, latency, and communication overhead.}

For future work, it would be interesting to extend our results to the online and dynamic setting, where the set of transactions to be processed is continuously generated and injected into a shard. Additionally, exploring efficient communication mechanisms between shards and considering the impact of network congestion, where network links have bounded capacity, would be interesting areas for further research.

\section*{Acknowledgments}
This paper is supported by NSF grant 2131538.

\bibliographystyle{IEEEtran}
\bibliography{references}

\begin{thebibliography}{10}
\providecommand{\url}[1]{#1}
\csname url@samestyle\endcsname
\providecommand{\newblock}{\relax}
\providecommand{\bibinfo}[2]{#2}
\providecommand{\BIBentrySTDinterwordspacing}{\spaceskip=0pt\relax}
\providecommand{\BIBentryALTinterwordstretchfactor}{4}
\providecommand{\BIBentryALTinterwordspacing}{\spaceskip=\fontdimen2\font plus
\BIBentryALTinterwordstretchfactor\fontdimen3\font minus \fontdimen4\font\relax}
\providecommand{\BIBforeignlanguage}[2]{{%
\expandafter\ifx\csname l@#1\endcsname\relax
\typeout{** WARNING: IEEEtran.bst: No hyphenation pattern has been}%
\typeout{** loaded for the language `#1'. Using the pattern for}%
\typeout{** the default language instead.}%
\else
\language=\csname l@#1\endcsname
\fi
#2}}
\providecommand{\BIBdecl}{\relax}
\BIBdecl

\bibitem{survey-of-onsensus}
L.~S. Sankar, M.~Sindhu, and M.~Sethumadhavan, ``Survey of consensus protocols on blockchain applications,'' in \emph{2017 4th international conference on advanced computing and communication systems (ICACCS)}.\hskip 1em plus 0.5em minus 0.4em\relax IEEE, 2017, pp. 1--5.

\bibitem{bitcoin}
S.~Nakamoto, ``Bitcoin : A peer-to-peer electronic cash system,'' 2009.

\bibitem{ethereum}
V.~Buterin \emph{et~al.}, ``A next-generation smart contract and decentralized application platform,'' \emph{white paper}, vol.~3, no.~37, pp. 2--1, 2014.

\bibitem{mcghin2019blockchain}
T.~McGhin, K.-K.~R. Choo, C.~Z. Liu, and D.~He, ``Blockchain in healthcare applications: Research challenges and opportunities,'' \emph{Journal of network and computer applications}, vol. 135, pp. 62--75, 2019.

\bibitem{10575641}
R.~L. Neupane, E.~Bonnah, B.~Bhusal, K.~Neupane, K.~A. Hoque, and P.~Calyam, ``Formal verification for blockchain-based insurance claims processing,'' in \emph{NOMS 2024-2024 IEEE Network Operations and Management Symposium}, 2024, pp. 1--5.

\bibitem{akbarfam2023forensiblock}
A.~J. Akbarfam, M.~Heidaripour, H.~Maleki, G.~Dorai, and G.~Agrawal, ``Forensiblock: A provenance-driven blockchain framework for data forensics and auditability,'' \emph{arXiv preprint arXiv:2308.03927}, 2023.

\bibitem{akbarfam2023deep}
A.~J. Akbarfam, S.~Barazandeh, D.~Gupta, and H.~Maleki, ``Deep learning meets blockchain for automated and secure access control,'' \emph{arXiv preprint arXiv:2311.06236}, 2023.

\bibitem{azzi2019power}
R.~Azzi, R.~K. Chamoun, and M.~Sokhn, ``The power of a blockchain-based supply chain,'' \emph{Computers \& industrial engineering}, vol. 135, pp. 582--592, 2019.

\bibitem{al2021scichain}
A.~Al-Mamun, F.~Yan, and D.~Zhao, ``Scichain: Blockchain-enabled lightweight and efficient data provenance for reproducible scientific computing,'' in \emph{2021 IEEE 37th International Conference on Data Engineering (ICDE)}.\hskip 1em plus 0.5em minus 0.4em\relax IEEE, 2021, pp. 1853--1858.

\bibitem{9430722}
Z.~Ning, S.~Sun, X.~Wang, L.~Guo, S.~Guo, X.~Hu, B.~Hu, and R.~Y.~K. Kwok, ``Blockchain-enabled intelligent transportation systems: A distributed crowdsensing framework,'' \emph{IEEE Transactions on Mobile Computing}, vol.~21, no.~12, pp. 4201--4217, 2022.

\bibitem{9387145}
J.~Li, J.~Wu, L.~Chen, J.~Li, and S.-K. Lam, ``Blockchain-based secure key management for mobile edge computing,'' \emph{IEEE Transactions on Mobile Computing}, vol.~22, no.~1, pp. 100--114, 2023.

\bibitem{yuan2022coopedge}
L.~Yuan, Q.~He, S.~Tan, B.~Li, J.~Yu, F.~Chen, and Y.~Yang, ``Coopedge+: Enabling decentralized, secure and cooperative multi-access edge computing based on blockchain,'' \emph{IEEE Transactions on Parallel and Distributed Systems}, vol.~34, no.~3, pp. 894--908, 2022.

\bibitem{cocco2017banking}
L.~Cocco, A.~Pinna, and M.~Marchesi, ``Banking on blockchain: Costs savings thanks to the blockchain technology,'' \emph{Future internet}, vol.~9, no.~3, p.~25, 2017.

\bibitem{chen2020blockchain}
Y.~Chen and C.~Bellavitis, ``Blockchain disruption and decentralized finance: The rise of decentralized business models,'' \emph{Journal of Business Venturing Insights}, vol.~13, p. e00151, 2020.

\bibitem{adhikari2023lockless}
\BIBentryALTinterwordspacing
R.~Adhikari and C.~Busch, ``Lockless blockchain sharding with multiversion control,'' in \emph{Structural Information and Communication Complexity: 30th International Colloquium, SIROCCO 2023, Alcal\'{a} de Henares, Spain, June 6–9, 2023, Proceedings}.\hskip 1em plus 0.5em minus 0.4em\relax Berlin, Heidelberg: Springer-Verlag, 2023, p. 112–131. [Online]. Available: \url{https://doi.org/10.1007/978-3-031-32733-9_6}
\BIBentrySTDinterwordspacing

\bibitem{adhikari2024spaastable}
R.~Adhikari, C.~Busch, and D.~Kowalski, ``Stable blockchain sharding under adversarial transaction generation,'' in \emph{Proceedings of the 36th ACM Symposium on Parallelism in Algorithms and Architectures}, 2024.

\bibitem{Elastico}
L.~Luu, V.~Narayanan, C.~Zheng, K.~Baweja, S.~Gilbert, and P.~Saxena, ``A secure sharding protocol for open blockchains,'' in \emph{Proceedings of the 2016 ACM SIGSAC conference on computer and communications security}, 2016, pp. 17--30.

\bibitem{OmniLedger}
E.~Kokoris-Kogias, P.~Jovanovic, L.~Gasser, N.~Gailly, E.~Syta, and B.~Ford, ``Omniledger: A secure, scale-out, decentralized ledger via sharding,'' in \emph{2018 IEEE Symposium on Security and Privacy (SP)}.\hskip 1em plus 0.5em minus 0.4em\relax IEEE, 2018, pp. 583--598.

\bibitem{Rapidchain}
\BIBentryALTinterwordspacing
M.~Zamani, M.~Movahedi, and M.~Raykova, ``Rapidchain: Scaling blockchain via full sharding,'' in \emph{Proceedings of the 2018 ACM SIGSAC Conference on Computer and Communications Security}, ser. CCS '18.\hskip 1em plus 0.5em minus 0.4em\relax New York, NY, USA: Association for Computing Machinery, 2018, p. 931–948. [Online]. Available: \url{https://doi.org/10.1145/3243734.3243853}
\BIBentrySTDinterwordspacing

\bibitem{amiri2021sharper}
M.~J. Amiri, D.~Agrawal, and A.~El~Abbadi, ``Sharper: Sharding permissioned blockchains over network clusters,'' in \emph{Proceedings of the 2021 international conference on management of data}, 2021, pp. 76--88.

\bibitem{xu2024x}
J.~Xu, Y.~Ming, Z.~Wu, C.~Wang, and X.~Jia, ``X-shard: Optimistic cross-shard transaction processing for sharding-based blockchains,'' \emph{IEEE Transactions on Parallel and Distributed Systems}, 2024.

\bibitem{Byshard}
J.~Hellings and M.~Sadoghi, ``Byshard: Sharding in a byzantine environment,'' \emph{Proceedings of the VLDB Endowment}, vol.~14, no.~11, pp. 2230--2243, 2021.

\bibitem{de2018impact}
G.~De~Masi, ``The impact of topology on internet of things: A multidisciplinary review,'' in \emph{2018 Advances in Science and Engineering Technology International Conferences (ASET)}.\hskip 1em plus 0.5em minus 0.4em\relax IEEE, 2018, pp. 1--6.

\bibitem{mamat2019network}
H.~Mamat, B.~Ibrahim, and M.~Sulong, ``Network topology comparison for internet communication and iot connectivity,'' in \emph{2019 IEEE Conference on Open Systems (ICOS)}.\hskip 1em plus 0.5em minus 0.4em\relax IEEE, 2019, pp. 1--5.

\bibitem{PBFT}
M.~Castro, B.~Liskov \emph{et~al.}, ``Practical byzantine fault tolerance,'' in \emph{OsDI}, vol.~99, 1999, pp. 173--186.

\bibitem{gupta2006oblivious}
A.~Gupta, M.~T. Hajiaghayi, and H.~R{\"a}cke, ``Oblivious network design,'' in \emph{Proceedings of the seventeenth annual ACM-SIAM symposium on Discrete algorithm}, 2006, pp. 970--979.

\bibitem{yin2019hotstuff}
M.~Yin, D.~Malkhi, M.~K. Reiter, G.~G. Gueta, and I.~Abraham, ``Hotstuff: Bft consensus with linearity and responsiveness,'' in \emph{Proceedings of the 2019 ACM Symposium on Principles of Distributed Computing}, 2019, pp. 347--356.

\bibitem{Jalal-Window}
M.~M. {Jalalzai} and C.~{Busch}, ``Window based {BFT} blockchain consensus,'' in \emph{iThings, IEEE GreenCom, IEEE (CPSCom) and IEEE SSmartData 2018}, July 2018, pp. 971--979.

\bibitem{jalalzai2019proteus}
M.~M. Jalalzai, C.~Busch, and G.~G. Richard, ``Proteus: A scalable bft consensus protocol for blockchains,'' in \emph{2019 IEEE international conference on Blockchain (Blockchain)}.\hskip 1em plus 0.5em minus 0.4em\relax IEEE, 2019, pp. 308--313.

\bibitem{jalalzai2021hermes}
M.~M. Jalalzai, C.~Feng, C.~Busch, G.~G. Richard, and J.~Niu, ``The hermes bft for blockchains,'' \emph{IEEE Transactions on Dependable and Secure Computing}, vol.~19, no.~6, pp. 3971--3986, 2021.

\bibitem{stathakopoulou2019mir}
C.~Stathakopoulou, T.~David, and M.~Vukolic, ``Mir-bft: High-throughput bft for blockchains,'' \emph{arXiv preprint arXiv:1906.05552}, vol.~92, 2019.

\bibitem{spiegelman2022bullshark}
A.~Spiegelman, N.~Giridharan, A.~Sonnino, and L.~Kokoris-Kogias, ``Bullshark: Dag bft protocols made practical,'' in \emph{Proceedings of the 2022 ACM SIGSAC Conference on Computer and Communications Security}, 2022, pp. 2705--2718.

\bibitem{danezis2022narwhal}
G.~Danezis, L.~Kokoris-Kogias, A.~Sonnino, and A.~Spiegelman, ``Narwhal and tusk: a dag-based mempool and efficient bft consensus,'' in \emph{Proceedings of the Seventeenth European Conference on Computer Systems}, 2022, pp. 34--50.

\bibitem{dang2019towards}
H.~Dang, T.~T.~A. Dinh, D.~Loghin, E.-C. Chang, Q.~Lin, and B.~C. Ooi, ``Towards scaling blockchain systems via sharding,'' in \emph{Proceedings of the 2019 international conference on management of data}, 2019, pp. 123--140.

\bibitem{li2023lb}
M.~Li, W.~Wang, and J.~Zhang, ``Lb-chain: Load-balanced and low-latency blockchain sharding via account migration,'' \emph{IEEE Transactions on Parallel and Distributed Systems}, vol.~34, no.~10, pp. 2797--2810, 2023.

\bibitem{queralta2021blockchain}
J.~P. Queralta and T.~Westerlund, ``Blockchain for mobile edge computing: Consensus mechanisms and scalability,'' \emph{Mobile edge computing}, pp. 333--357, 2021.

\bibitem{li2021blockchain}
G.~Li, X.~Ren, J.~Wu, W.~Ji, H.~Yu, J.~Cao, and R.~Wang, ``Blockchain-based mobile edge computing system,'' \emph{Information Sciences}, vol. 561, pp. 70--80, 2021.

\bibitem{9324984}
X.~Cai, S.~Geng, J.~Zhang, D.~Wu, Z.~Cui, W.~Zhang, and J.~Chen, ``A sharding scheme-based many-objective optimization algorithm for enhancing security in blockchain-enabled industrial internet of things,'' \emph{IEEE Transactions on Industrial Informatics}, vol.~17, no.~11, pp. 7650--7658, 2021.

\bibitem{ren2023data}
Y.~Ren, X.~Liu, P.~K. Sharma, O.~Alfarraj, A.~Tolba, S.~Wang, and J.~Wang, ``Data storage mechanism of industrial iot based on lrc sharding blockchain,'' \emph{Scientific Reports}, vol.~13, no.~1, p. 2746, 2023.

\bibitem{set2022service}
S.~K. Set and G.~S. Park, ``Service-aware dynamic sharding approach for scalable blockchain,'' \emph{IEEE Transactions on Services Computing}, 2022.

\bibitem{busch2023stable}
C.~Busch, B.~S. Chlebus, D.~R. Kowalski, and P.~Poudel, ``Stable scheduling in transactional memory,'' in \emph{Algorithms and Complexity: 13th International Conference, CIAC 2023, Larnaca, Cyprus, June 13--16, 2023, Proceedings}.\hskip 1em plus 0.5em minus 0.4em\relax Springer, 2023, pp. 172--186.

\bibitem{attiya2015directory}
H.~Attiya, V.~Gramoli, and A.~Milani, ``Directory protocols for distributed transactional memory,'' \emph{Transactional Memory. Foundations, Algorithms, Tools, and Applications: COST Action Euro-TM IC1001}, pp. 367--391, 2015.

\bibitem{sharma2014distributed}
G.~Sharma and C.~Busch, ``Distributed transactional memory for general networks,'' \emph{Distributed computing}, vol.~27, no.~5, pp. 329--362, 2014.

\bibitem{sharma2015load}
------, ``A load balanced directory for distributed shared memory objects,'' \emph{Journal of Parallel and Distributed Computing}, vol.~78, pp. 6--24, 2015.

\bibitem{busch2017fast}
C.~Busch, M.~Herlihy, M.~Popovic, and G.~Sharma, ``Fast scheduling in distributed transactional memory,'' in \emph{Proceedings of the 29th ACM Symposium on Parallelism in Algorithms and Architectures}, 2017, pp. 173--182.

\bibitem{busch2022dynamic}
------, ``Dynamic scheduling in distributed transactional memory,'' \emph{Distributed Computing}, vol.~35, no.~1, pp. 19--36, 2022.

\bibitem{hellings2022fault}
J.~Hellings and M.~Sadoghi, ``The fault-tolerant cluster-sending problem,'' in \emph{Foundations of Information and Knowledge Systems: 12th International Symposium, FoIKS 2022, Helsinki, Finland, June 20--23, 2022, Proceedings}.\hskip 1em plus 0.5em minus 0.4em\relax Springer, 2022, pp. 168--186.

\bibitem{leighton2014introduction}
F.~T. Leighton, \emph{Introduction to parallel algorithms and architectures: Arrays{\textperiodcentered} trees{\textperiodcentered} hypercubes}.\hskip 1em plus 0.5em minus 0.4em\relax Elsevier, 2014.

\bibitem{chan1989embedding}
M.-Y. Chan, ``Embedding of d-dimensional grids into optimal hypercubes,'' in \emph{Proceedings of the first annual ACM symposium on Parallel algorithms and architectures}, 1989, pp. 52--57.

\bibitem{li2021scaling}
M.~Li and Y.~Qin, ``Scaling the blockchain-based access control framework for iot via sharding,'' in \emph{ICC 2021-IEEE International Conference on Communications}.\hskip 1em plus 0.5em minus 0.4em\relax IEEE, 2021, pp. 1--6.

\end{thebibliography}


 


\vspace{11pt}


\begin{IEEEbiography}[{\includegraphics[width=1in,height=1.25in,clip,keepaspectratio]{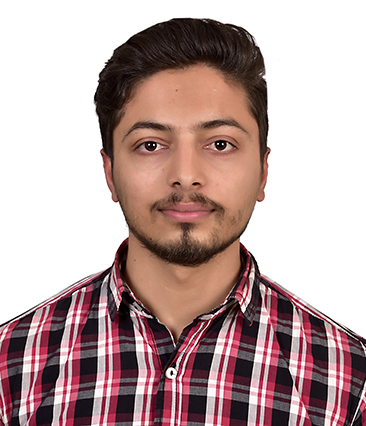}}]{Ramesh Adhikari}
received his B.E. in Computer Engineering from Tribhuvan University, Kathmandu, Nepal in 2017 and M.E. in Computer Engineering from Pokhara University, Nepal. He is currently working toward a PhD degree in the School of Computer and Cyber Sciences at Augusta University, USA. His research interests include blockchain and distributed computing.
\end{IEEEbiography}

\begin{IEEEbiography}[{\includegraphics[width=1in,height=1.25in,clip,keepaspectratio]{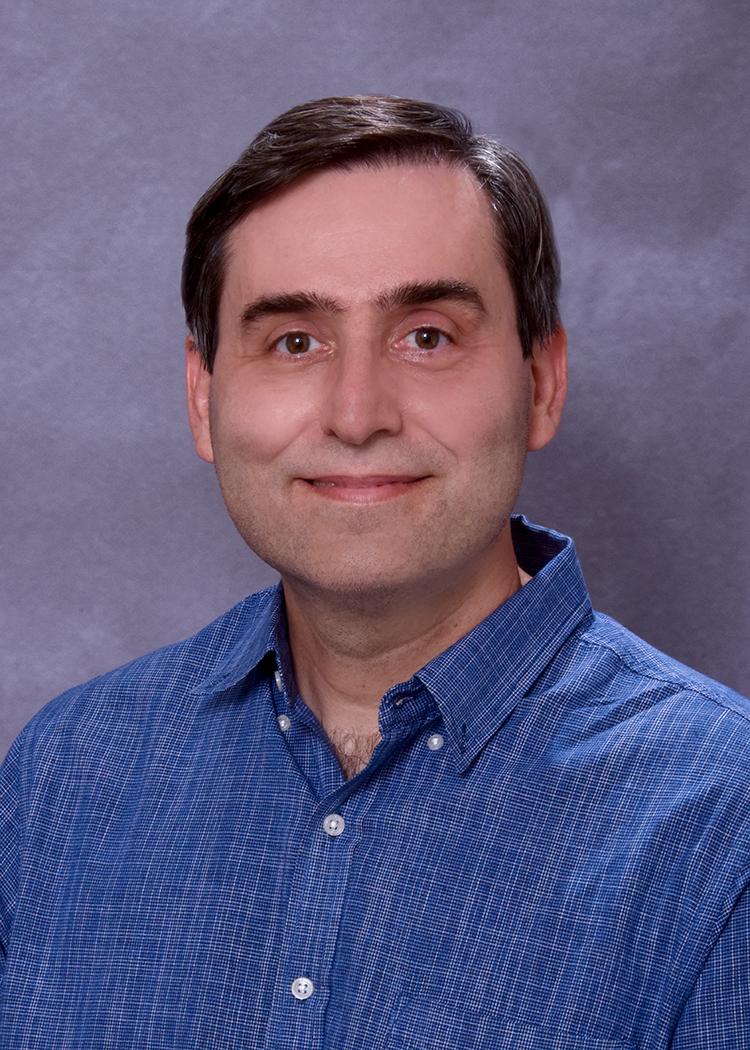}}]{Costas Busch}
obtained a B.Sc. in 1992 and a M.Sc. in 1995 in Computer Science from the University of Crete, Heraklion, Greece. He received a Ph.D. degree in Computer Science in 2000 from Brown University. Since 2020 he is a professor in the School of Computer and Cyber Sciences at Augusta University. His research interests are in distributed algorithms/data structures. He has over 130 publications in venues such as (IEEE) FOCS, IPDPS, and (ACM) JACM, STOC, PODC, SPAA. 
\end{IEEEbiography}


\begin{IEEEbiography}[{\includegraphics[width=1in,height=1.25in,clip,keepaspectratio]{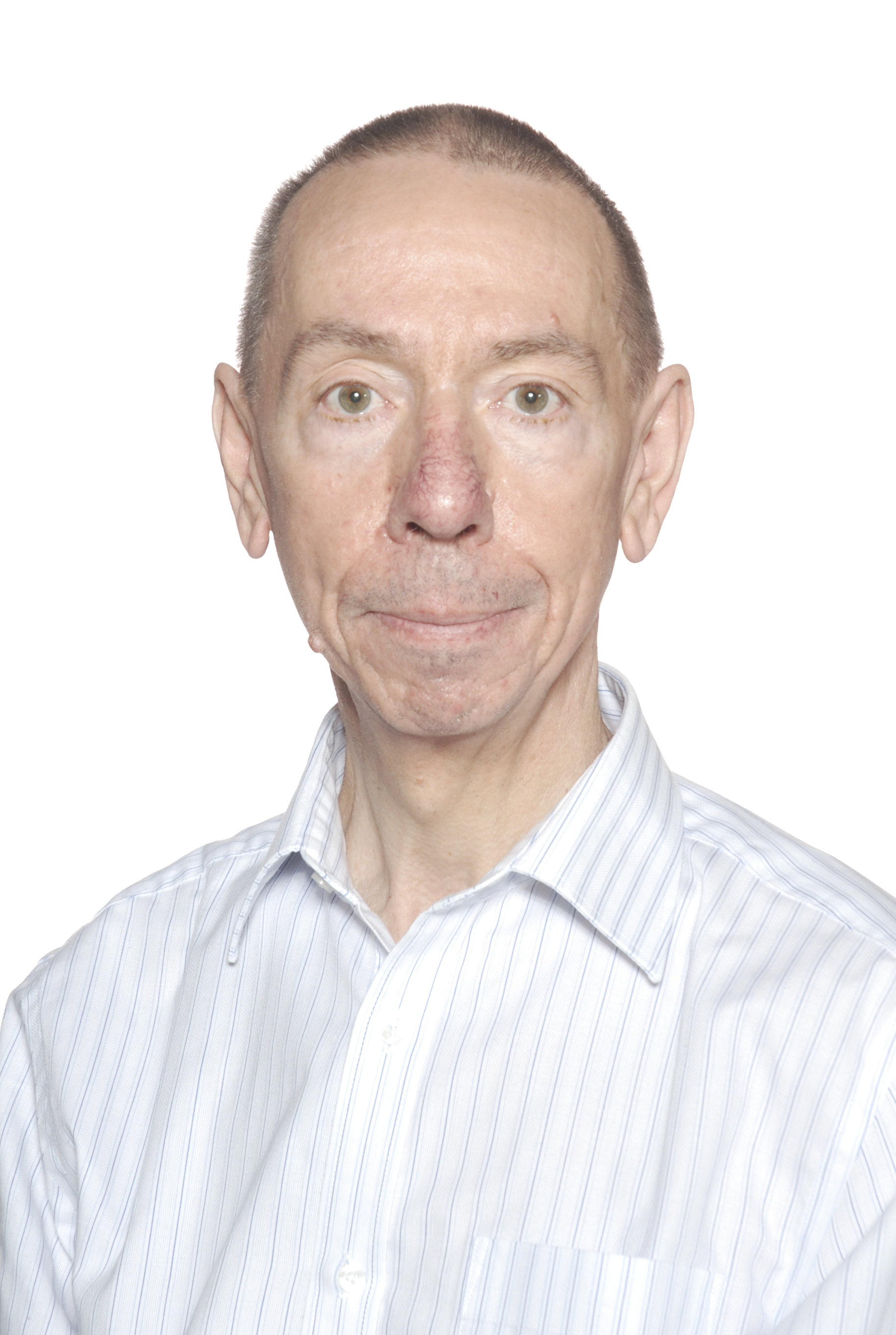}}]{Miroslav Popovic}
 received his Dipl. Eng., M.Sc., and Ph.D.  degrees from the Faculty of Technical Sciences, University of Novi Sad, Serbia, in 1984, 1988, and 1990, respectively. He is a Full Professor at the University of Novi Sad from 2002. He has authored or co-authored 33 peer-reviewed journal papers, more than 120 conference papers, and 5 patents. He is also the author of the internationally known book “Communication protocol engineering, Second Edition” (CRC Press, 2018). His research interests are engineering of computer-based systems, intelligent distributed systems, and security.
\end{IEEEbiography}

\vfill

\end{document}